%% file: usenix.tex
\documentclass[letterpaper,twocolumn,10pt]{article}
\usepackage{usenix,epsfig,endnotes}
\usepackage{ulem}
\usepackage{pifont}
\usepackage{xcolor}
\usepackage{tcolorbox}
\usepackage{amsmath}
\usepackage{enumitem}
\usepackage{array} 
\usepackage[caption=false,font=normalsize]{subfig}
\usepackage{appendix}
\usepackage{authblk}
\usepackage{ulem}
\usepackage{multirow}
\usepackage{makecell}
\usepackage{tcolorbox}
\usepackage{algorithm}
\usepackage{algpseudocode}
\usepackage{amsmath}
\usepackage[utf8]{inputenc}
\usepackage[T1]{fontenc} 
\usepackage{amssymb}    
\usepackage{booktabs}  
\usepackage{tabularx}
\usepackage{graphicx} 
\newcommand{\starRating}[1]{%
  \ifnum#1=1 \star \else
  \ifnum#1=2 \star\star \else
  \ifnum#1=3 \star\star\star \else
  \ifnum#1=4 \star\star\star\star \else
  \ifnum#1=5 \star\star\star\star\star \fi\fi\fi\fi
}
\usepackage{multirow}

\newcommand{\zyz}[1]{ \textcolor{red}{#1}}
\newcommand{\zyk}[1]{ \textcolor{blue}{#1}}

\newcolumntype{S}{>{\Large}c}

\newcommand{\projname}{\textsc{X\text{-}PRINT}}
\long\def\comment#1{}
\begin{document}

\date{}

\title{\projname{}: Platform-Agnostic and Scalable Fine-Grained Encrypted\\Traffic Fingerprinting}
\author[1]{YuKun Zhu}
\author[1]{ManYuan Hua}
\author[1]{Hai Huang}
\author[1]{YongZhao Zhang \textsuperscript{*}}
\author[1]{Jie Yang}
\author[1]{FengHua Xu}
\author[1]{RuiDong Chen}
\author[1]{XiaoSong Zhang \textsuperscript{*}}
\author[1]{JiGuo Yu}
\author[1]{Yong Ma}
\affil[1]{University of Electronic Science and Technology of China}


\maketitle

\thispagestyle{empty}

\begin{abstract}
Although encryption protocols such as TLS are widely deployed, side-channel metadata in encrypted traffic still reveals patterns that allow application and behavior inference. However, existing fine-grained fingerprinting approaches face two key limitations: (i) reliance on platform-dependent characteristics, which restricts generalization across heterogeneous platforms, and (ii) poor scalability for fine-grained behavior identification in open-world settings.

In this paper, we present \projname{}, the first server-centric, URI-based framework for cross-platform fine-grained encrypted-traffic fingerprinting. \projname{} systematically demonstrates that backend URI invocation patterns can serve as platform-agnostic invariants and are effective for modeling fine-grained behaviors. To achieve robust identification, \projname{} further leverages temporally structured URI maps for behavior inference and emphasizes the exclusion of platform- or application-specific private URIs to handle unseen cases, thereby improving reliability in open-world and cross-platform settings. Extensive experiments across diverse cross-platform and open-world settings show that \projname{} achieves state-of-the-art accuracy in fine-grained fingerprinting and exhibits strong scalability and robustness.


\end{abstract}
\comment{
Although TLS is widely used to protect data in transit, side-channel metadata still enables inference of sensitive user behaviors. Existing behavioral fingerprinting methods often depend on client-side instrumentation, limiting scalability across heterogeneous platforms, and struggle with the vast diversity of applications and frequent updates that introduce unseen targets. We present \projname{}, a server-side framework for cross-platform fine-grained behavior identification, modeling user activities as logically structured API request sequences. \projname{} applies platform-agnostic URI-level labeling, constructs stable URI Maps to capture invariant backend invocation patterns, and leverages shared URIs to transfer knowledge to unseen applications. Evaluations on encrypted traffic from multiple real-world platforms show that \projname{} reduces labeling noise by 42.7\%, improves the F1-score from 0.85 to 0.99, and maintains over 95\% cross-platform accuracy. Compared with FOAP and APPScanner, \projname{} achieves superior performance in distinguishing similar behaviors and platform variants while lowering false positives, demonstrating practical value for encrypted traffic analysis.
}

\input{1_intro.tex}

\input{2_background.tex}

\input{4_FlowLabel.tex}

\input{3_threat.tex}

\input{5_approach.tex}

\input{6_eval.tex}

\input{3_related.tex}
\input{8_conclusion.tex}
\cleardoublepage

{
\cleardoublepage
\normalem
\bibliographystyle{plain}
\bibliography{sample}

}

\cleardoublepage
\appendix



\section{Application Version}\label{sec:appendix_version}
To evaluate cross-version robustness, we collected traffic at two time points, October 2024 and August 2025, which are separated by an interval of eleven months. The dataset covers ten real-world applications from three categories (Video, Social, and Music), and for each application, we paired the client version observed in October~2024 with the version available in August~2025 to form an old$\rightarrow$new comparison used throughout the analysis.
\begin{table}[!ht]
\caption{List of Application Versions.}
\vspace{1em}
\begin{tabular}{llll}
\toprule
\textbf{Category} & \textbf{App}           & \textbf{Old Ver.}  & \textbf{New Ver.}  \\ \midrule
Video    & YouTube       & 19.30.36  & 20.29.39  \\
         & Twitch        & 20.6.0    & 25.4.0    \\
         & TED           & 7.5.47    & 7.5.74    \\ \hline
Social   & Weibo         & 14.10.3   & 15.8.0    \\
         & X             & 10.53.0   & 11.10.0   \\
         & Reddit        & 2024.31.0 & 2025.30.0 \\
         & Bluesky       & 1.91.2    & 1.105.0   \\ \hline
Music    & YouTube Music & 7.21.50   & 8.30.51   \\
         & PocketFM      & 6.5.2     & 8.6.5     \\ 
         & QQ Music      &  13.1.0.8 & 14.8.0.3  \\ \bottomrule
\end{tabular}
\end{table}

Inspection of the version numbers shows heterogeneous evolution. Several applications underwent major upgrades, while others received only incremental maintenance releases during the eleven-month window. This pattern is consistent with our recognition results: major upgrades are more likely to alter behavior signatures such as URI templates and invocation order, thereby increasing the chance of cross version mismatches; minor releases tend to preserve behavior-level characteristics, so their impact on identification accuracy is comparatively smaller.

\section{URI Map-based Matching Algorithm}
\label{sec:map-matching}

We adopt a URI map-based matching algorithm (Algorithm~\ref{alg:WeightedLCS}) to estimate the likelihood that a predicted URI map (e.g., $\mathcal{S}$) corresponds to a candidate behavior’s CUM (e.g., $\mathcal{C}$).  We fix the confidence gate $\tau=0.5$. For each candidate behavior $b$ with domain-partitioned CUM $\{T_b^{(d)}\}$, we build the unique URI set $\mathcal{C}=\bigcup_{d}\mathrm{set}\!\big(T_b^{(d)}\big)$. From the predicted sequence $\mathcal{S}=\{(s_i,p_i)\}_{i=1}^m$, we define the per-URI maximum confidence $p_u=\max\{\,p_i:\ s_i=u\,\}$ and the covered set $\mathcal{P}=\mathcal{C}\cap\{\,u:\ p_u\ \text{is defined}\,\}$. For each domain $d$, we run $\textsc{LCSMatch}(\mathcal{S}^{(d)},T_b^{(d)};\tau)$ under the URI-map predicate with the gate $\tau$, which returns a monotone matching $\mathcal{M}^{(d)}$; its domain weight is $w^{(d)}=\sum_{u_i\in\mathcal{M}^{(d)}} p_{u_i}$ using only terms with $p_{u_i}\ge 0.5$. The numerator aggregates across domains as $w^{(d)}=\sum_{u_i\in\mathcal{M}} p_{u_i}$ where $\mathcal{M}=\mathcal{M} \cup \mathcal{M}^{(d)}$. The denominator is $\sum_{u_j\in\mathcal{P}} p_{u_j}+\lambda\big(|\mathcal{C}|-|\mathcal{P}|\big)$. The resulting score is $\mathrm{Score}_{\text{map}}(b)=w^{(d)}/\big(\sum_{u_j\in\mathcal{P}} p_{u_j}+\lambda(|\mathcal{C}|-|\mathcal{P}|)\big)$ , and the predicted label is $\hat b=\arg\max_{b\in\mathcal{B}}\mathrm{Score}_{\text{map}}(b)$. Algorithm~\ref{alg:WeightedLCS} implements this procedure.

\input{Table/BehaviorMap}

\section{Impact of Cross-Dataset Evaluation}
In this experiment, we evaluate the impact of the sample data automatically collected by \projname{} from scripts on the identification of real human operations.

\input{Table/RHBehavior.tex}\noindent\textbf{Experimental setup.} To assess generalization from the randomized traces, we additionally built a human-operated dataset: five volunteers manually interacted with 100 applications, producing 500 traffic instances with realistic usage patterns on the same three platforms. We construct a human generated dataset by randomly sampling 100 applications from a pool of 1,000 and recruiting five volunteers. Each volunteer operates all 100 apps for about five minutes per session, producing 5$\times$100 traffic instances. We evaluate three transfer settings: A$\rightarrow$A trains and tests \projname{} on Airtest generated traffic from the same domain; A$\rightarrow$H trains on Airtest traffic and tests on human traffic; AH$\rightarrow$H augments the Airtest training set with human traffic from four volunteers and tests on the remaining volunteer. We rotate the held out volunteer and report the average for statistical reliability. Models are trained only on the designated source data and evaluated on the target without additional tuning. We report behavior level precision, recall, and F1.

\noindent \textbf{Results.} The results in Table~\ref{tab:RHbehavior} show high identification accuracy in \textit{A$\rightarrow$A} and \textit{A$\rightarrow$H}, indicating that our Airtest based collection effectively reproduces human behavior. In \textit{A+H$\rightarrow$H}, \projname{} attains perfect precision, recall, and F1. Compared with FOAP, \projname{} maintains stable performance, whereas FOAP’s F1 drops in \textit{A$\rightarrow$H} due to context dependent inference. This experiment highlights an efficient training data strategy that combines automated scripts for large scale collection with human sessions to capture real world operational logic, thereby reducing false negatives and improving recall for \projname{}.

\end{document}

%% file: 1_intro.tex
\section{Introduction}

End-user devices, such as smartphones, laptops, and IoT devices have become indispensable in daily life~\cite{dumitrescu2014new}, with diverse applications greatly extending their functionality and reshaping modern lifestyles~\cite{odoom2022mobile}. Human actions have become deeply interwoven with the digital domain, wherein daily behaviors and social interactions are increasingly represented and recorded through networked services~\cite{bastos2021spatializing}.

Modern applications, while offering convenience, involve vast amounts of user data, raising significant privacy concerns~\cite{tangari2021mobile,liu2022protecting}. Despite the widespread use of encrypted communication, they still remain vulnerable to traffic fingerprinting, including app fingerprinting (AF) attacks~\cite{2016AppScanner,taylor2017robust} and fine-grained behavioral fingerprinting (BF) attacks~\cite{ma2020pinpointing,prasad2023context}, where adversaries identify applications or detailed application behaviors by exploiting distinctive traffic patterns without accessing packet payloads. 
However, existing fine-grained traffic fingerprinting approaches are often platform-dependent, facing two major limitations that hinder their effectiveness in real-world cross-platform scenarios.

\textbf{Platform-Dependent Characteristics.} Modern network environments comprise interleaved traffic generated by millions of heterogeneous devices running diverse operating systems (e.g., Android, iOS, Windows, and OpenWRT), each hosting applications with distinct implementations and runtime conditions~\cite{naboulsi2015large}. Existing fine-grained behavior identification techniques often rely on client-side instrumentation at varying granularities—session-level~\cite{velan2015survey}, activity-level~\cite{heng2021utmobilenettraffic2021}, and UI-level~\cite{li2022foap}—and are therefore inherently platform dependent. These discrepancies, compounded by differences in operating systems, device configurations, and background services, make it difficult to obtain a consistent representation of behavior across platforms. \textit{Therefore, there is a pressing need for platform-agnostic characteristics that remain applicable in heterogeneous environments.}

\textbf{Scalability and Identification Granularity.}
For fine-grained traffic fingerprinting, prior work needs to build profiles for each application, platform, and software version (for both OS versions and application versions), which is not scalable in cross-platform and open-world settings. This is because the same application across platforms or versions often undergoes substantial distribution shifts, yielding noticeable changes in the feature space.
\textit{Therefore, it is crucial to extract invariant traffic patterns to support scalable fine-grained traffic fingerprinting for unseen cases}, 
which can largely improve our understanding of real-world traffic landscape.

In this paper, we present \projname{}, a novel server-side, URI-based framework for fine-grained and scalable encrypted traffic fingerprinting across heterogeneous platforms that addresses the above limitations. First, unlike client-side instrumentation, \projname{} proposes a server-centric perspective that treats backend URI (Uniform Resource Identifier) invocation patterns as platform-agnostic characteristics. The core insight is that cross-platform clients commonly rely on shared backend services operated by overlapping providers; for example, every YouTube client ultimately communicates with Google's backend to retrieve video content. By elevating backend URIs to the unit of analysis, this server-centric view decouples recognition from client UIs, OS idiosyncrasies, and instrumentation choices. To validate this insight, we deploy 
a controlled Man-in-the-Middle (MitM) environment and directly inspect URI invocations. We observe that (i) different client platforms invoke a substantially overlapping set of network URIs, ranging from around 30\% to nearly 90\%, and (ii) the resulting URI-request sequences are sufficiently stable and distinctive to support fine-grained traffic fingerprinting.

Second, \projname{} implements a URI-based framework for real-world encrypted traffic inference. It trains lightweight classifiers in the local MitM environment to identify backend URIs from side-channel features of flows (e.g., packet sizes, directions, and timestamps), and the resulting models can be applied directly to real-world encrypted traffic without assuming MitM access. Moreover, to improve the robustness with the presence of interleaved traffic, \projname{} adopts a two-stage pipeline for fine-grained encrypted traffic fingerprinting: (i) Coarse-grained app filtering based on flow-level features, where we do not attempt full app identification but instead filter out irrelevant applications to shrink the search space under cross-platform complexity; and (ii) Fine-grained behavior matching based on URI Maps, where we determine the application and specific behavior by aligning predicted URI sequences to pre-built URI maps constructed from training data. The URI map based matching integrates both the temporal structure of URI invocation patterns and the confidence scores of individual URI predictions, enabling effective discrimination among similar behaviors across diverse apps.



Third, in cross-platform and open-world settings where many applications are remaining unseen, \projname{} can still infer fine-grained behaviors for these apps. We focus on unseen cases that pass both the coarse-grained filtering and fine-grained matching stages—such as functionally similar applications, deployments of the same application on previously unobserved platforms, or major version updates; by contrast, meaningful inference for entirely unrelated applications is impractical (but can be removed). Our key insight is that customized (platform- or app-specific) URI invocations—those absent from the training set—introduce substantial noise into the matching process. Accordingly, when a case is flagged as unseen (after the second-stage fine-grained matching), we refine the matching to the subset of shared URIs and disregard private URIs. Importantly, this procedure does not require estimating the exact proportion of shared URIs; it suffices to suppress noise by excluding private URIs.

Our evaluations show that \projname{} improves the F1-score of fine-grained behavior identification by 45\% over state-of-the-art baselines with interleaved network traffic from multiple platforms and end-users.
In addition, \projname{} can accurately detect the presence of unseen cases in open-world scenarios. Finally, by applying a refinement strategy that excludes application- or platform-specific private features from unseen cases, \projname{} further achieves accurate and robust fine-grained inference for unseen applications, platforms, and versions. This capability is largely absent from prior work, demonstrating strong scalability of \projname{}.


Our contributions are summarized as follows:

\begin{itemize}
\item To the best of our knowledge, we are the first to propose a server-centric, URI-based framework for cross-platform fine-grained traffic fingerprinting, revealing the potential for backend URI invocation patterns to serve as platform-agnostic characteristics.


\item We design a two-stage framework for fine-grained encrypted traffic inference based on temporal-structured URI maps. We also identify the importance to distinguish shared and private URIs in open-world settings.


\item We implement an \projname{} prototype and validate its performance across cross-platform and open-world settings, where it significantly outperforms baseline approaches in fine-grained behavior recognition.


\end{itemize}

%% file: 2_background.tex
\section{Background}\label{sec:background}
  

\subsection{Basics of Traffic Encryption}
Modern Internet services are typically built upon structured request protocols such as HTTP and gRPC, where each request targets a specific Uniform Resource Identifier (URI, a standardized identifier for network resources) that identifies backend functionality. To safeguard the content of these requests, end-to-end encryption protocols such as TLS and QUIC are applied, encapsulating payloads and preventing intermediate observers from accessing application-layer data. Nevertheless, each encrypted packet still discloses side-channel metadata, including its \textit{packet size}, \textit{transmission direction} (client-to-server or server-to-client), and precise \textit{timestamp}~\cite{zhang2022defeating}. Prior studies have shown these side-channel information to be useful for various analytical tasks, including fraud detection~~\cite{alwhbi2024encrypted}, anomaly detection~\cite{sattar2025anomaly}, quality-of-service tuning~\cite{zhao2024metarocketc}, and forensic investigation~\cite{xu2022seeing}. These findings confirm that encrypted traffic continues to leak meaningful app fingerprints. However, accurately inferring fine-grained user actions still remains challenging due to the noisy, interleaved, and ambiguous nature of real-world traffic.



\subsection{Client-Side Traffic Identification}

Early approaches to behavior identification  from encrypted traffic relied on coarse-grained timestamp annotations, where all packets within the selected window following a user interaction were labeled with that action~\cite{draper2016characterization, sharafaldin2018toward, 2016AppScanner}. Such coarse temporal labeling ignores network semantics and cannot separate user-initiated foreground activities from overlapping background services (e.g., advertising SDKs, system telemetry, OS heartbeats), thereby introducing substantial noise. Network dynamics, including latency fluctuations, retransmissions, and packet reordering, further blur behavior boundaries and degrade model performance.

A recent work, FOAP~\cite{li2022foap}, improved upon this by instrumenting Android applications to intercept activity lifecycle events and UI callbacks, aligning each traffic burst with a precise user action. While this strategy filters out much of the background noise, it is inherently tied to client-side instrumentation and specific platform implementations. As a result, although FOAP achieves fine-grained user action identification on Android, its reliance on platform-dependent characteristics limits scalability to heterogeneous platforms.




\subsection{Interleaved Cross-Platform Traffic}
In real-world environments, network traffic from heterogeneous platforms such as Android, iOS, and desktop clients frequently coexists. Additionally, multiple users often share the same network, resulting in highly interleaved traffic flows. These flows are end-to-end encrypted, which obscures client-specific metadata and complicates behavior inference. Heterogeneous system architectures further exacerbate the challenge, as client-side instrumentation is difficult to generalize across platforms (e.g., FOAP is limited to Android). Considering the large number of applications, directly adapting Android-based instrumentation to platforms such as iOS, PCs, or IoT devices is impractical, requiring labor-intensive alignment. This difficulty arises because the generated labels are inconsistent across platforms due to divergent UI workflows, SDK logic, and runtime scheduling policies. As a result, in realistic network environments, encrypted traffic from multiple platforms and end-users is often interleaved,
\textbf{it is essential to identify platform-agnostic characteristics of encrypted traffic that remain applicable across heterogeneous systems.}

%% file: 4_FlowLabel.tex
\section{URI-based Server-Side Traffic Analysis}\label{sec:flowlabel}
This section presents a preliminary study to analyze the usage patterns of server-side URI invocations across different platforms and applications.



\subsection{Motivation}



Our goal is to infer fine-grained user actions from encrypted traffic generated by different platforms (e.g., Android, iOS, Windows, etc.), and even include unseen platforms and applications during the training phase. 
The key insight is that \textit{different platform versions of the same application typically share a common backend run by the same provider} (e.g., YouTube clients talk to Google and selected third-party services). These backends expose server-side URI requests that deliver or collect data (video content, search results, user profiles), while invocation sequences encode the underlying business logic of the traffic. Therefore, rather than analyzing client-specific features as in prior work~\cite{draper2016characterization,lin2022bert,ren2019international}, we take a server-centric view and use URI-request patterns as platform-independent characteristics for fine-grained behavior identification.

To validate this idea, we need to answer the following two questions.  \textbf{Q1:} Do different client platforms indeed invoke an overlapping subset of network URIs? If so, those shared endpoints can act as anchors across platforms, enabling us to recognize the same behavior even when the traffic originates from unseen platforms.  \textbf{Q2:} Are the resulting URI-request sequences sufficiently stable and distinguishable to support fine-grained behavior identification?



\comment{
Achieving fine-grained and cross-platform user behavior recognition from encrypted traffic is particularly challenging because it requires distinguishing semantically precise user actions while ensuring robustness across heterogeneous client platforms. Prior methods typically rely on client-side instrumentation or statistical features extracted from encrypted flows, both of which suffer from platform-specific variability and are vulnerable to pervasive background noise from SDKs, system processes, and ad libraries. These limitations obscure the semantic structure of user behaviors and undermine the stability and generalizability of inference models, especially in cross-platform scenarios. A promising direction arises from a key observation: despite differences in client-side implementations, many applications are supported by a unified server-side service provider that exposes standardized APIs for delivering core functionalities such as search, playback, and login. This consistency suggests that the server-side logic triggered by user actions tends to remain stable across platforms and manifests as discernible and structured API invocation patterns. These patterns are less affected by platform-dependent noise compared to client-side signals and may be observable even in encrypted traffic. This observation raises two essential questions that determine the feasibility of our approach. \textbf{Q1:} Can shared server-side APIs serve as semantic anchors for identifying platform-independent behaviors, while private APIs reflect platform-specific variations? \textbf{Q2:} Are API invocation sequences sufficiently stable to support robust and interpretable behavior modeling? These questions represent necessary conditions for building a server-centric analysis framework in encrypted traffic scenarios.
}

\comment{

\zyz{While client-side execution may vary due to UI pipelines, SDK dependencies, or OS-specific scheduling, the server-side logic invoked by user actions remains highly consistent.} 

Key observation: API invoke

Question 1: shared API.

Question 2: pattern?

Despite differences in client-side implementations, these platforms are typically supported by a \textit{unified backend service provider}, which delivers a consistent set of functionalities through standardized server-side APIs. \zyz{?} This architectural convergence ensures that semantically equivalent user behaviors-such as search, playback, or login-are implemented through identical API endpoints across platforms.\zyz{?}

\zyk{The presence of a \zyz{shared} backend infrastructure plays a pivotal role in enabling cross-platform behavior modeling. In particular, \zyz{\textit{shared APIs? Question 1}} encode the core service semantics and are reused across platforms to ensure functional uniformity and simplify backend maintenance.}

\zyk{From a traffic analysis perspective, this consistent reuse of server APIs offers a unique advantage: although encrypted traffic conceals payload contents, \zyz{the patterns of API invocation? Question 2}-especially those corresponding to shared APIs-remain observable and structurally stable. These patterns can serve as reliable and interpretable anchors for modeling user behaviors in a platform-agnostic manner.}

\zyz{Why our approach works} \zyk{However, prior approaches often focus on client-side instrumentation or rely on statistical features extracted from encrypted flows, both of which are sensitive to platform-specific variability and lack semantic clarity. This motivates our server-centric approach, which leverages shared API invocation sequences as a foundation for fine-grained, cross-platform behavior recognition under encryption.} \zyz{?}
x
}


\comment{
\subsection{Overview}

This section articulates and validates the central hypothesis of this work: 
\textit{server-side API invocation patterns serve as robust and platform-independent semantic anchors for fine-grained behavior identification, even under encryption constraints}.

Traditional encrypted traffic analysis approaches often rely on statistical features or deep learning representations of raw packet sequences. However, these methods lack interpretability and fail to generalize across heterogeneous platforms due to high variability in packet-level signals. Moreover, they ignore the intrinsic semantic structure of user behavior as reflected in application-layer interactions.

To address these limitations, we shift the focus to API-level semantics. Our core assumption is that user actions consistently trigger a small set of API calls with stable temporal and structural properties, regardless of encryption or platform differences. These API sequences implicitly encode user intent and can be exploited as high-level features for cross-platform behavior recognition.

To substantiate this view, we organize this section around three key questions:
\begin{itemize}
    \item \textbf{Q1:} Is it feasible to extract API-level semantics from encrypted traffic via practical means such as MITM interception?
    \item \textbf{Q2:} Do these extracted APIs demonstrate consistent structural and temporal characteristics across platforms?
    \item \textbf{Q3:} Can the sequence and combination of API calls reliably reflect user behavior, enabling accurate cross-platform inference?
\end{itemize}

Through empirical analysis, we demonstrate that API patterns offer a viable and interpretable foundation for behavior modeling under encryption. These findings directly motivate the design of the proposed \projname{} system, introduced in the next section.
}

\subsection{URI Extraction}\label{sec:uri_extraction}
To analyze the URI invocations for different applications and platforms, we deploy a controlled MitM environment~\cite{chordiya2018man} in which a client device runs the target application while its traffic is transparently decrypted, enabling us to extract the full URI requests, parameters, and transmission metadata. MitM is a common technique for intercepting and analyzing encrypted traffic by acting as an intermediary between the client and server~\cite{conti2016survey}. Specifically, we deploy a proxy based on mitmproxy~\cite{kim2021mitm}, and use Frida~\cite{pourali2024racing} scripts to bypass certificate validation, enabling traffic decryption across Android, iOS, and Windows.

\begin{figure}[t]
    \centering
    \subfloat[Two flows triggered by the same user action.\label{fig:packet2API}]{\includegraphics[width=\linewidth]{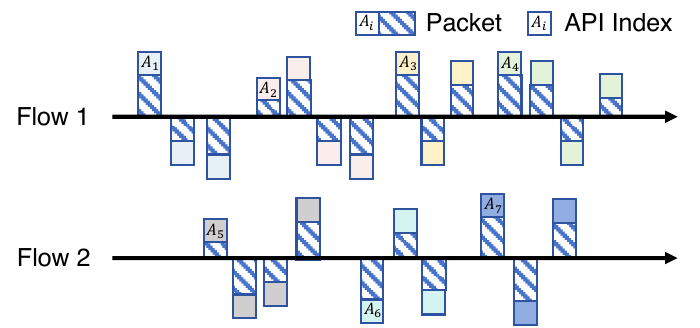}}
    \hfill
    \centering
    \subfloat[Dividing flow 1 into shared and private parts according to the URI labels of each packet.\label{fig:API_Separation}]{\includegraphics[width=\linewidth]{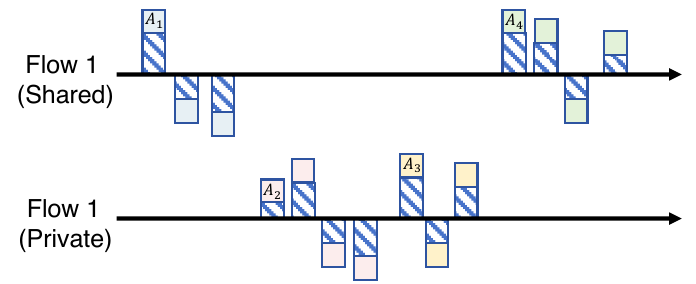}}
    \hfill
    \subfloat[URI map constructed by two flows.\label{fig:API_Map}]{\includegraphics[width=0.9\linewidth]{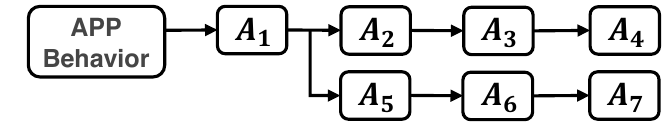}}
    \caption{Demonstrating the URI-based traffic analysis. (a) We can inspect the URI label of each packet in the controlled MitM environment. (b) Encrypted packets can be grouped according to the shared and private URIs. (c) The URI invocation sequence captures the response logic of backend servers.}
\end{figure}

An example URI triggered by the YouTube `Search' behavior observed in the decrypted traffic is: \texttt{POST https://www.\\youtube.com/youtubei/v1/search?key=...}, where URI path \texttt{https://www.youtube.com/youtubei/v1/search} (we use \texttt{/search} for short hereafter) denotes the backend interface being accessed, and the accompanying parameters (e.g., search query, device context, language settings) denote user- or platform-specific input. While parameters may vary across each invocation to deliver contextual details, the URI path remains unchanged and reflects to backend functionality. For each encrypted packet, we can extract its associated URI information as shown in Fig.~\ref{fig:packet2API}, and the timestamp of each packet reflects the order of URI requests. As a result, we collect decrypted traffic from several applications including Reddit, RedNote, YouTube, and Twitch across Android, iOS, and Windows platforms, and each collected packet is labeled with its corresponding URI path, making it possible to analyze the usage patterns of network URIs.

\subsection{Cross-Platform URI Usage Analysis}\label{sec:uri_usage}






\subsubsection{Cross-Platform Consistency}\label{sec:uri_consistency}

We first try to investigate whether URI requests exhibit consistent invocation patterns across different platforms (\textbf{Q1}).

\textbf{Shared URIs V.S. Private URIs.}
For each application, we enumerate every URI invoked by the `Search' behavior across three platforms (e.g., Android, iOS, and Windows) and classify them into two categories: \textit{Shared URIs} and \textit{Private URIs}. Shared URIs are those that appear across the three platforms and expose identical or structurally equivalent URI paths, regardless of parameter variations. For instance, \texttt{/search} is consistently used for search functionality across all YouTube clients. In contrast, Private URIs are observed exclusively on a single platform, typically supporting auxiliary features, device-specific optimizations, or UI-level logic.

Table~\ref{tab:Public} summarizes the number of observed URIs and the invocation frequency of shared URIs (the proportion of packets labeled with shared URIs regarding the behavior `Search') across different platforms. We can draw three observations: (i) A proportion of URIs are shared across platforms, indicating that many applications rely on a common backend infrastructure to deliver core functionalities. (ii) The usage frequency of shared URIs varies across applications and platforms, from 30\% to 89\% in our dataset, suggesting that the encrypted traffic contains substantial shared features. (iii) The private URIs may be caused by device-specific UI elements, auxiliary services, or custom backend logic, can contribute to platform-specific features to enable the platform identification.



\input{Table/apitotal.tex}

\textbf{Similarities of the Encrypted Flow Patterns.}
To further investigate the existence of cross-platform patterns introduced by shared URIs, we divide the packets of each encrypted flow into two parts, i.e., the packets corresponding to shared URIs and those corresponding to private URIs, as shown in Fig.~\ref{fig:API_Separation}. We then compute the similarity of the encrypted flow patterns between different platforms based on the DTW distance~\cite{tao2021comparative}, which is a widely used method for measuring the similarity of time series data. 
The average similarities across applications and platforms are shown in Table~\ref{tab:Cross-Platform}. The packets corresponding to shared URIs achieve much higher similarity scores, with an average of approximately 0.86 over all platform pairs, while that of private URIs only yield much lower scores, often below 0.25. This consistency suggests that it is possible to leverage the shared URIs for cross-platform identification based on the encrypted traffic data.

\input{Table/apianalysis.tex}

\subsubsection{URI Maps Triggered by User Behaviors}\label{sec:uri-map}

Fig.~\ref{fig:API_Map} illustrates that a single application behavior typically triggers multiple URI requests (consecutive packets with the same URI label are grouped together), due to the involvement of multiple flows connecting to various back-end services such as content retrieval, logging, or personalization.
Our next question is whether such sequential information is stable enough to support more robust behavior recognition (\textbf{Q2})? 
Table~\ref{tab:cbs_top3} lists the top-3 URI sequences observed for a single flow triggered by the behaviors on YouTube of each platform; the subscript attached to every URI denotes its rank in the overall frequency list (rank 1 means the most frequent). We only show the URIs regarding Google servers for simplicity, while numerous third-party URIs are discarded. We can draw the following observations: 

\textbf{Canonical URI Map (CUM).} 
For each application behavior, e.g., Android/Search, the sequential order of URI requests may vary due to network conditions. However, one sequence tends to appear most frequently, which we refer to as the \textit{Canonical URI Map} (CUM). For example, in Android/Search, the CUM is [$A_3$, $A_5$, $A_9$, $A_{12}$, $A_{27}$, $A_{16}$], with a proportion of around 76\%. Moreover, though the URI map slightly varies, certain key URIs repeatedly appear in each sequence, such as $A_3$ (\path{/v1}), $A_{12}$ (\path{/guide}), $A_{27}$ (\path{/history}), and $A_{16}$ (\path{/search}). Such URIs must be invoked to complete the intended behavior, e.g., $A_{16}$ is always invoked when the user initiates a search action. Therefore, leveraging the CUM of each behavior enables us to construct a more robust identification scheme.

\textbf{Discrepancy Across Behaviors.}
For two behaviors in the same platform, like Android/Search and Android/Play, the URI maps may share some common invocations, such as $A_3$ and $A_{5}$. However, $A_{16}$ (\texttt{/search}) and $A_{30}$ (\texttt{/playback}) differs the URI maps accordingly. As a result, the URI maps can be used to distinguish fine-grained application behaviors since each behavior should request different contents from the server, thereby invoking different URIs.



\textbf{Cross-Platform Generalization.}
For the same behavior across different platforms, such as `Search' on Android, iOS, and Web, the CUM significantly differs due to the platform-specific implementations. However, the \texttt{/search} URI is consistently invoked in all three platforms, indicating the same semantic intent of the behavior (e.g., issuing requests to Google servers for search results) across platforms. As a result, if we can identify the key URIs, such as $A_{16}$, we can infer that the user is possibly performing a search action on YouTube, even if the traffic is generated by an unseen platform, while the rest of the URI requests help us to further refine the behavior identification and distinguish between platforms.


\input{Table/sequence.tex}


%% file: Table/apitotal.tex
\begin{table}[t]
\centering
\small
\caption{Shared URIs across  different platforms.}
\vspace{0.5em} 
\label{tab:Public}
\begin{tabular}{ccccc}
\toprule
\textbf{Application} & \textbf{Platform} & \textbf{\makecell[c]{Total\\URIs}} & \textbf{\makecell[c]{Shared\\URIs}} & \textbf{\makecell[c]{Invocation\\Frequency}} \\
\midrule
\multirow{3}{*}{Reddit}
    & Android & 21 & \multirow{3}{*}{12} & 0.303 \\
    & iOS     & 27 &                     & 0.360\\
    & Windows     & 28 &                     & 0.541 \\
\midrule
\multirow{3}{*}{RedNote} 
    & Android & 94  & \multirow{3}{*}{15} & 0.471 \\
    & iOS     & 101 &                     & 0.287 \\
    & Windows     & 48  &                     & 0.410 \\
\midrule
\multirow{3}{*}{YouTube} 
    & Android & 31 & \multirow{3}{*}{13} & 0.483 \\
    & iOS     & 39 &                     & 0.427 \\
    & Windows     & 49 &                     & 0.426 \\
\midrule
\multirow{3}{*}{Twitch}
    & Android & 33 & \multirow{3}{*}{11} & 0.807 \\
    & iOS     & 16 &                     & 0.896 \\
    & Windows     & 35 &                     & 0.589 \\
\bottomrule
\end{tabular}
\end{table}

%% file: Table/apianalysis.tex
\begin{table}[t]
\centering
\small 
\caption{Similarity of encrypted traffic.}
\vspace{0.5em}
\label{tab:Cross-Platform}
\begin{tabular}{ccc}
\toprule
\textbf{Traffic Type} & \textbf{Platform Pair} & \textbf{Avg. Similarity} \\
\midrule
\multirow{3}{*}{Shared URIs}
    & Android vs. iOS & 0.865 \\
    & Android vs. Windows  & 0.855 \\
    & iOS vs. Windows      & 0.864 \\
\midrule
\multirow{3}{*}{Private URIs}
    & Android vs. iOS & 0.162 \\
    & Android vs. Windows  & 0.239 \\
    & iOS vs. Windows      & 0.205 \\
\bottomrule
\end{tabular}
\end{table}


%% file: Table/sequence.tex
\begin{table}[t]
\centering
\small 
\caption{Statistics of the top-3 URI invocation sequences.}
\vspace{0.5em} 
\label{tab:cbs_top3}
\resizebox{\linewidth}{!}{
\begin{tabular}{SlS}
\toprule
\textbf{Platform/Behavior} & \textbf{\Large{\makecell[c]{URI Map}}} & \textbf{Proportion} \\
\midrule
\multirow{3}{*}{Android/Search} 
               & [$A_3$, $A_5$, $A_9$, $A_{12}$, $A_{27}$, $A_{16}$] & 0.76 \\
               & [$A_3$, $A_5$, $A_9$, $A_{27}$, $A_{16}$] & 0.10 \\
               & [$A_3$, $A_5$, $A_{12}$, $A_{27}$, $A_{16}$] & 0.07 \\
\midrule 
\multirow{3}{*}{Android/Play} 
               & [$A_3$, $A_5$, $A_9$, $A_{12}$, $A_{30}$] & 0.78 \\
               & [$A_3$, $A_5$, $A_9$, $A_{12}$, $A_{30}$, $A_8$] & 0.10 \\
               & [$A_3$, $A_5$, $A_{12}$, $A_{30}$] & 0.06 \\
\midrule
\multirow{3}{*}{iOS/Search}     
               & [$A_3$, $A_5$, $A_8$, $A_{16}$] & 0.68 \\
               & [$A_3$, $A_{12}$, $A_8$, $A_8$, $A_9$, $A_{20}$, $A_5$, $A_{20}$, $A_5$, $A_{16}$] & 0.24 \\
               & [$A_3$, $A_5$, $A_{16}$] & 0.06 \\
\midrule
\multirow{3}{*}{Windows/Search}     
               & [$A_{12}$, $A_{20}$, $A_{22}$, $A_{23}$, $A_{25}$, $A_{23}$, $A_{20}$, $A_{12}$, $A_{16}$] & 0.59 \\
               & [$A_{12}$, $A_{20}$, $A_{22}$, $A_{25}$, $A_{23}$, $A_{20}$, $A_{12}$, $A_{16}$] & 0.10 \\
               & [$A_{12}$, $A_{23}$, $A_{20}$, $A_{22}$, $A_{25}$, $A_{23}$, $A_{20}$, $A_{12}$, $A_{16}$] & 0.09 \\
\bottomrule
\end{tabular}
}
\end{table}


%% file: 3_threat.tex
\section{Threat Model}\label{sec:threatmodel}
\begin{figure}[t]
    \centering
    \includegraphics[width=\linewidth]{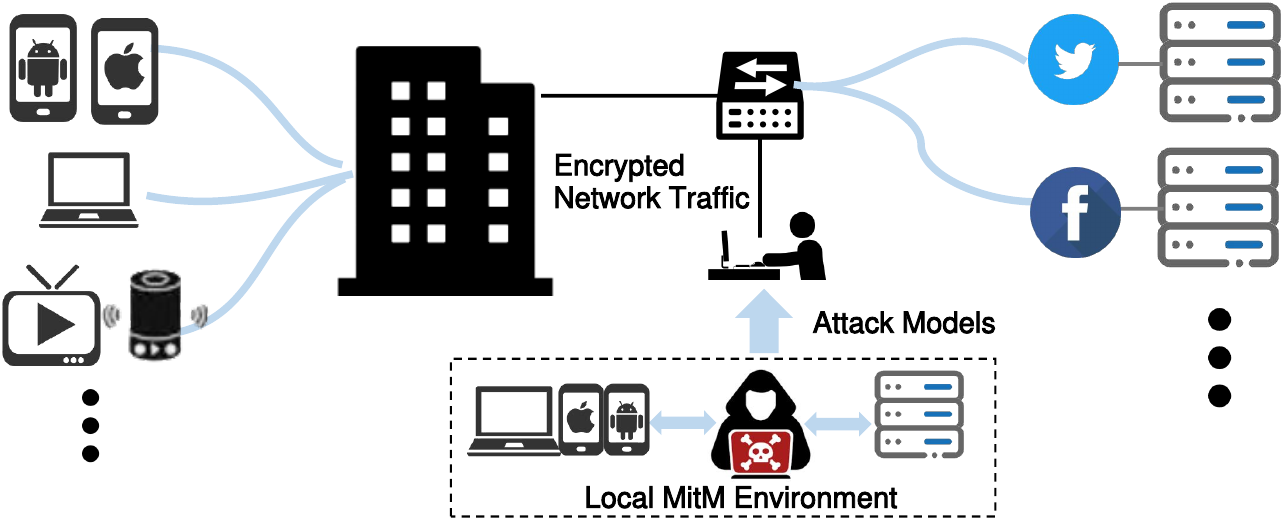}
    \caption{The threat model of \projname{}.}
    \label{threat model}
\end{figure}

The attacker's goal is to infer fine-grained application behaviors at the central switch or gateway by passively monitoring encrypted traffic. Note that, as shown in Figure~\ref{threat model}, there could be multiple applications of interest and heterogeneous platforms (e.g., Android, iOS, Windows, and IoT devices, etc.) interleaved in network traffic. Therefore, we establish the following assumptions for the attacker to achieve this goal:

\textbf{Controlled Local MitM Environment for Model Training:} The attacker is capable of setting up a controlled local MitM environment, \textit{on their own devices}, to collect encrypted traffic traces and decrypt URIs for offline analysis and model training (e.g., the models to infer URIs from encrypted traffic data). Note that, such plaintext traces accessibility is not assumed in the real-world attack scenario.

\textbf{Encrypted Traffic Observation for Inference:} The attacker can only observe encrypted traffic for real-world implementation, which is protected by protocols such as TLS or QUIC. This means that in the inference stage, the attacker cannot access the packet payloads or any application-specific identifiers (e.g., URIs) that may be present in the plaintext.

\textbf{No Proximity Access to the Victim Devices:} The attacker does not have direct access to the victim devices, such as smartphones, tablets, or PCs. This means they cannot install malicious software or directly compromise the victim devices.


%% file: 5_approach.tex

\section{\projname{} Design}\label{sec:design}
In this section, we discuss how to implement \projname{} without MitM access in real attack scenarios.




\begin{figure*}[ht]
\centering
\includegraphics[width=1\linewidth]{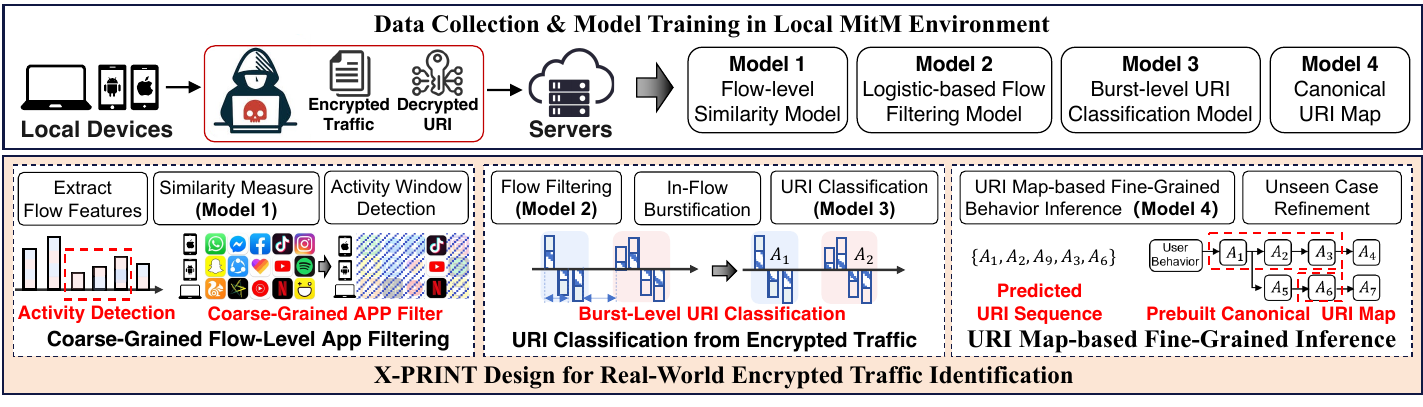}
\vspace{-15pt}
\caption{
Overview of \projname{}. In the controlled local MitM environment, attackers collect and analyze network traffic across different platforms using their own devices, training four models for real-world inference. Specifically, \projname{} leverages flow-level features for coarse-grained app filtering and burst-level URIs as the analytical unit for fine-grained behavior inference.
}

\label{fig:workflow}
\end{figure*}

\subsection{System Overview}\label{sec:workflow}
As shown in Figure~\ref{fig:workflow}, \projname{} consists of the following four main modules.

\noindent\textbf{$\bullet$ Decrypted Traffic Collection \& Model Training in Controlled Local Environment (Section~\ref{sec:uri_extraction}).}
We established a controlled MitM environment to collect encrypted traffic and annotate flows with URI-level labels. Specifically, this step produces four types of models for real world inference:
(i) flow similarity models, (ii) logistic models for flow filtering, (iii) URI classification models, and (iv) CUMs.

\noindent\textbf{$\bullet$ Coarse-Grained Flow-Level App Filtering (Section~\ref{sec:coarse}).}  
For real world implementation, \projname{} first extracts side-channel features to compute flow-level similarity scores with known applications, before detecting activity windows and filtering our irrelevant segments. This step ensures that only traffic with sufficient similarity is preserved, reducing noise and computational overhead for fine-grained inference.

\noindent\textbf{$\bullet$ URI Classification from Encrypted Traffic (Section~\ref{sec:uri_classification}).}  
Next we perform in-flow burstification and classify bursts into URIs (rely on side-channel features of encrypted traffic), producing a sequence of predicted URIs. These URI sequences serve as the basis for the subsequent fine-grained inference.

\noindent\textbf{$\bullet$ URI Map-based Fine-Grained Inference (Section~\ref{sec:api_id}).} Finally, \projname{} compares the predicted URI sequences with the prebuilt CUMs to identify the detailed application behaviors. For the detected unseen cases, \projname{} further refines the matching process by excluding platform-specific URIs to improve the identification reliability.

In summary, the controlled MitM environment allows us to associate plaintext URI labels with their corresponding side-channel patterns in encrypted traffic. By learning mappings from side-channel features to URI labels, \projname{} can infer URI invocations in encrypted settings. Furthermore, the two-stage identification scheme, coupled with the refinement strategy for unseen cases, enhances accuracy and robustness in real-world deployment.



\subsection{Coarse-Grained Flow-Level App Filtering}\label{sec:coarse}

Our goal at this stage is to coarsely strip away the overwhelming majority of background and non-target traffic while preserving flows that may plausibly resemble traffic from the target application of interest, denoted as $\mathcal{A}$.



\subsubsection{Feature Extraction of Encrypted Traffic}\label{sec:flowFeature}


How to convert encrypted traffic flows into quantifiable measurements is crucial. For a given flow, no matter how long it is (e.g., a single flow, cascaded flows, or even in-flow bursts), we can extract a 123-dimensional feature vector from five perspectives~\cite{xu2022seeing,li2022packet,li2022foap}: general characteristics (8 features), including packet numbers, percentages, bytes, and duration; interactive patterns (20 features) using a function to characterize endpoint interactions; packet rate characteristics (5 features) based on packet duration divided into 1-second time windows; temporal characteristics (39 features), considering bidirectional, inbound, and outbound packet arrivals with various statistics on packet interval time and percentiles of relative arrival times; and packet size characteristics (51 features) covering various statistics on packet size, including mean, maximum, minimum, variance, skew, and percentiles for inbound, outbound, and bidirectional packets, etc.

\subsubsection{Activity Detection and Coarse-Grained Filtering
}
\label{sec:segments-filter}


Given the interleaved nature of traffic from multiple platforms and end-users in the complex real-world network environment, firstly, we need to identify the time windows where flows are likely triggered by the behaviors of application $\mathcal{A}$. However, the same applications running on different platforms and different applications with the similar functionalities (e.g., Music apps) may produce comparable encrypted traffic characteristics. Such cross-platform and cross-application similarities make it difficult to, at the flow-level, accurately attribute individual flows to their true sources. By contrast, it is possible to filter out the majority of background and irrelevant traffic while retaining flows that coarsely match the pattern of the target application $\mathcal{A}$. To achieve this goal, we design the following three-step pipeline to detect the activity windows.



\textbf{Similarity Measure:} This step first measures the similarity between a network flow and the traffic patterns of $\mathcal{A}$ in the feature space. We adopt a random-forest classifier in a one-vs-rest manner (i.e., computing the possibility that one flow belongs to the app of interest), as ensemble methods have demonstrated competitive accuracy and robustness in encrypted traffic classification in prior studies~\cite{anderson2017machine}.
Specifically, given a flow $f_i$, we extract its 123-dimensional feature vector $F(f_i)$ and feed it into the pre-trained model to obtain the application similarity score $p_i \in [0,1]$, where higher scores indicate closer resemblance to the target traffic patterns and lower scores suggest background or other applications. Note that, each application in the training set has its own model to compute the application similarity scores of given flows.



\textbf{Activity Window Detection:} Instead of directly applying the application similarity to determine whether a flow is triggered by user actions of app $\mathcal{A}$, we further leverage temporal context to reduce false positives.
Since a single user action typically triggers several flows whose similarity scores $p_i$ are strongly time-correlated. We therefore segment the traffic (with multiple flows) according to the sequential scores whenever the statistics change sharply. Segmentation follows two heuristics: (i) within a genuine activity window the variance of $p_i$ is low, whereas the variance between neighboring windows is high; and (ii) the total number of windows should be minimal, since consecutive high-score windows usually originate from the same application. We borrow the idea from classic voice-activity-detection (VAD) segmentation~\cite{patrona2016visual,ozturk2024radiovad,mariotte2024channel} tasks and treat the score series ${p_i}$ as an `energy' signal. Then we apply the Divisive-Agglomerative (D-A) Tree algorithm~\cite{roux2018comparative} to recursively split the traffic into segments.


\textbf{Coarse-Grained App Filtering for Activity Windows:} 
Since it is difficult to accurately determine whether an activity window belongs to the target application $\mathcal{A}$ based solely on flow-level features~\cite{ahmed2018statistical}, we adopt a relatively loose filtering criterion to coarsely select a set of potential applications that may generate these flows, and classify the true source of each flow in the fine-grained identification stage (Section.~\ref{sec:api_id}). Specifically, a segment is considered as potentially belonging to $\mathcal{A}$ if at least $q$\% of its flows have a similarity score $p_i > p_{\min}$, where $p_{\min}$ is the per-flow threshold and $q$ is the voting ratio. In our configuration, we empirically set $q = 0.8$ and $p_{\min} = 0.5$ to balance between retaining target-like segments and filtering out irrelevant traffic. Segments failing this criterion are discarded without further processing in this stage, thereby reducing the computational overhead of the cascaded fine-grained inference (i.e., Section~\ref{sec:uri_classification} and \ref{sec:api_id}). Note that, the filtering results determine the application-specific models to be used in the following tests.

\subsection{Burst-Level URI Classification}\label{sec:uri_classification}
Given multiple flows within a segment, in this section we try to classify each flow into the invoked server-side URIs.


\subsubsection{Fine-Grained Flow Filtering}\label{sec:irrelevant_flows}

The selected activity window still contain flows generated by background services or other applications, which may degrade the accuracy of subsequent URI-level inference. To further eliminate residual background flows, we adopt an additional background similarity score $r_i$, representing the probability that flow $f_i$ is pure background traffic. This score is obtained from a separate random-forest classifier trained on a dataset of representative background flows, including telemetry beacons, advertisement prefetches, and periodic push heartbeats, etc. Together with the application similarity score $p_i$ and the local temporal context $\bar{p}_{N_i}$, which denotes the average similarity of flows within a short time window around $f_i$, we construct a feature tuple $h_i = (p_i, \bar{p}_{N_i}, p_i - r_i)$. This tuple captures self-likeness, neighbor agreement, and app-versus-background contrast. A lightweight logistic-regression model evaluates each flow individually based on $h_i$ and outputs an acceptance probability. Flows whose acceptance probability exceeds the empirical gate threshold of 0.95 are retained for URI-level classification, whereas the others are discarded.

\subsubsection{In-Flow Burstification}

\begin{figure}[t]
    \centering
    \includegraphics[width=0.9\linewidth]{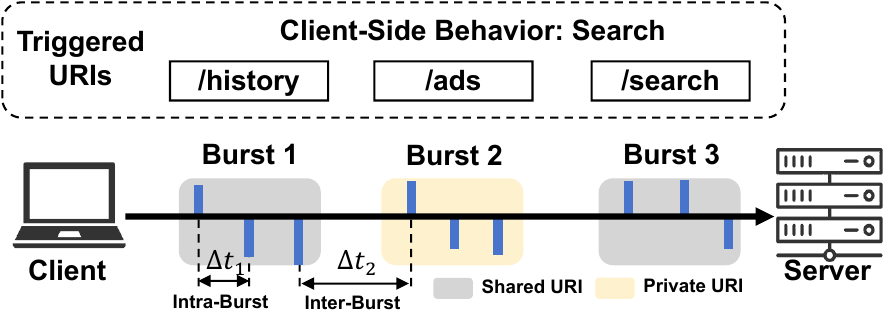}
    \caption{Flow burstification based on time intervals.}
    \label{fig:flowburst}
\end{figure}

Each flow may involve a sequence of URI invocations, as demonstrated in Fig.~\ref{fig:flowburst}.
Directly applying multi-label classification at the flow-level is impractical, as the number of URI combinations grows exponentially. 

To tackle this,
we partition each flow into packet clusters, namely \textit{bursts}~\cite{li2022foap}, before URI classification. 
As illustrated in Fig.~\ref{fig:flowburst}, a typical application behavior such as `Search' triggers multiple URI invocations (e.g., \texttt{/history}, \texttt{/ads}, \texttt{/search}, etc.), each corresponding to a distinct burst along the flow timeline. This is due to the fact that packets generated by the same URI request tend to appear in close temporal proximity, e.g., the \textit{Intra-Burst} interval ($\Delta t_1$) between two packets, while different URI requests are separated by significantly larger \textit{Inter-Burst} intervals ($\Delta t_2$). Such discrepancy can be simply captured by a predefined threshold $\epsilon$. This observation enables us to approximate URI request boundaries at the burst level, even for the encrypted traffic. 
Moreover, if a burst often contains two or more URIs due to the server-side design, we simply assign the label of the first observed URI in that burst. 


To implement the burstification strategy, specifically, we group consecutive packets into the same burst if the time interval between them is smaller than a threshold $\epsilon$; otherwise, a new burst is initiated. The value of $\epsilon$ plays a critical role in segmentation quality: smaller values may lead to over-burstification of a single URI invocation, while larger values may result in merging distinct invocations. We will evaluate the impact of different $\epsilon$ settings in Section~\ref{sec:effect_burst}. 
\begin{figure}[t]
    \centering
    \includegraphics[width=1\linewidth]{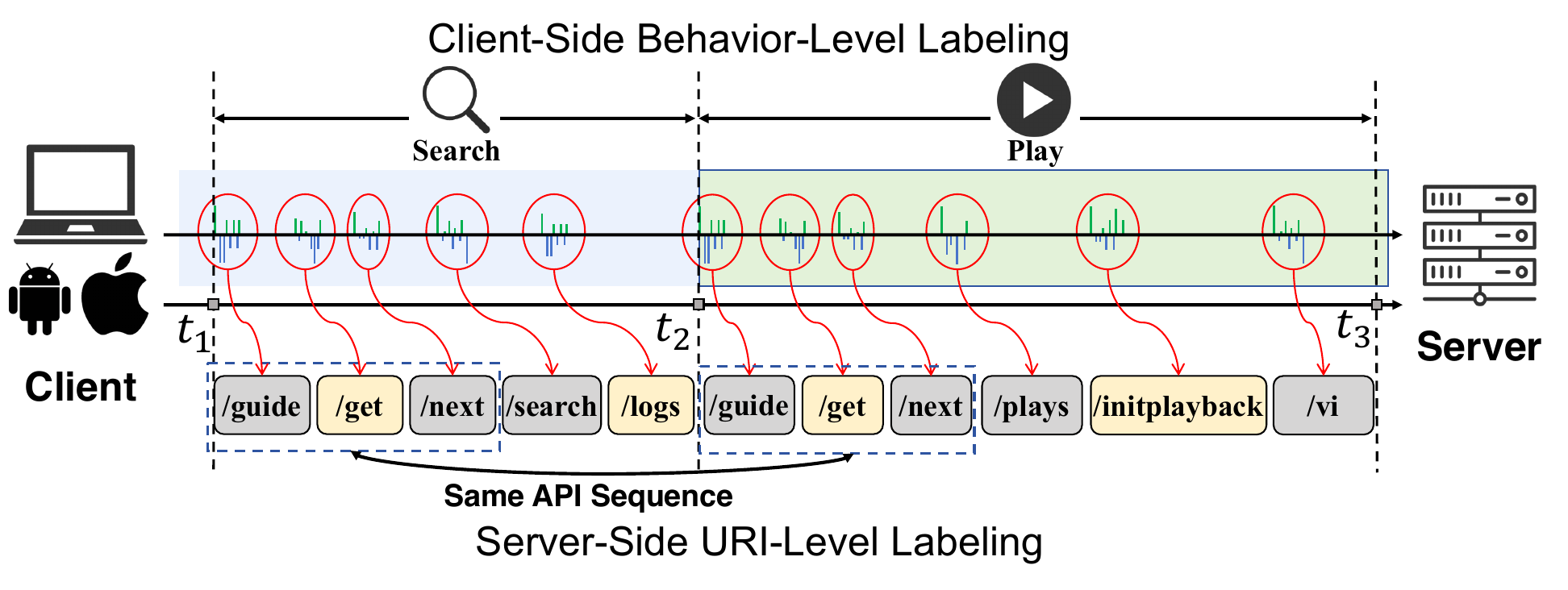}
    \vspace{-15pt}
    \caption{Comparison of client-side instrumentation for behavior-based labeling and server-side URI-based labeling.}
    \label{fig:Limitations-Burst}
\end{figure}

\subsubsection{Burst-based URI Classification}
While encryption prevents direct mapping between the content of individual packets and their server-side URIs, each URI invocation manifests as a burst of temporally correlated packets. These bursts exhibit distinguishable side-channel patterns (e.g., packet size, inter-arrival times), which allow us to reliably associate them with specific URIs.

Specifically, each burst can be represented by a statistical 123-dimensional feature vector (see Section~\ref{sec:flowFeature}) for classification. 
However, we found that the distribution of URIs is highly imbalanced. For example, URIs such as \texttt{/log} or \texttt{/ad} may occur frequently within a session, while core functional URIs (e.g., \texttt{/search}, \texttt{/play}) typically appear only once.
To mitigate this issue, we also adopt a random forest classifier for each application to perform URI classification, as this method has been widely recognized for its robustness and effectiveness in handling imbalanced classification tasks~\cite{anderson2017machine}.
The training data is obtained from decrypted traffic collected in the controlled MitM environment, where each burst is aligned with its ground-truth URI label via precise timestamp. 
For each flow in the inference stage, the output is an ordered URI sequence $\mathcal{S} = \{(s_1, p_1), (s_2, p_2), \dots, (s_m, p_m)\}$, where $s_j$ denotes the predicted server-side URI for the $j$-th burst and $p_j$ is its associated confidence score (from 0 to 1). 


\subsection{Fine-Grained Behavior Identification}\label{sec:api_id}

This section discusses how to conduct fine-grained behavior identification based on the burst-level URI sequences.

\subsubsection{Canonical URI Map Construction}

Different from existing client-side instrumentation approaches, URI invocation patterns are based on the server-response sequences, providing stable and semantically grounded representations of application behaviors across platforms.
As illustrated in Figure~\ref{fig:Limitations-Burst}, for instance, client-side instrumentation associates network flows with UI-level events~\cite{li2022foap}. For example, several bursts are associated with two user actions, `Search' and `Play', respectively, according to the client-side entry-point function invocation. However, this strategy introduces significant ambiguities: (i) nearly identical bursts (e.g., \texttt{/get}) may receive different labels due to UI variation, and (ii) distinct URI sequences (e.g., \texttt{/search} and \texttt{/logs}) may be assigned the same label despite invoking different backend services. In contrast, \projname{} classifies each burst into distinct server-side URIs based on their side-channel characteristics, thereby avoiding ambiguities introduced by client-side instrumentation and enabling more accurate and consistent behavior inference across platforms.

\begin{figure}[t]
    \centering
    \includegraphics[width=0.7\linewidth]{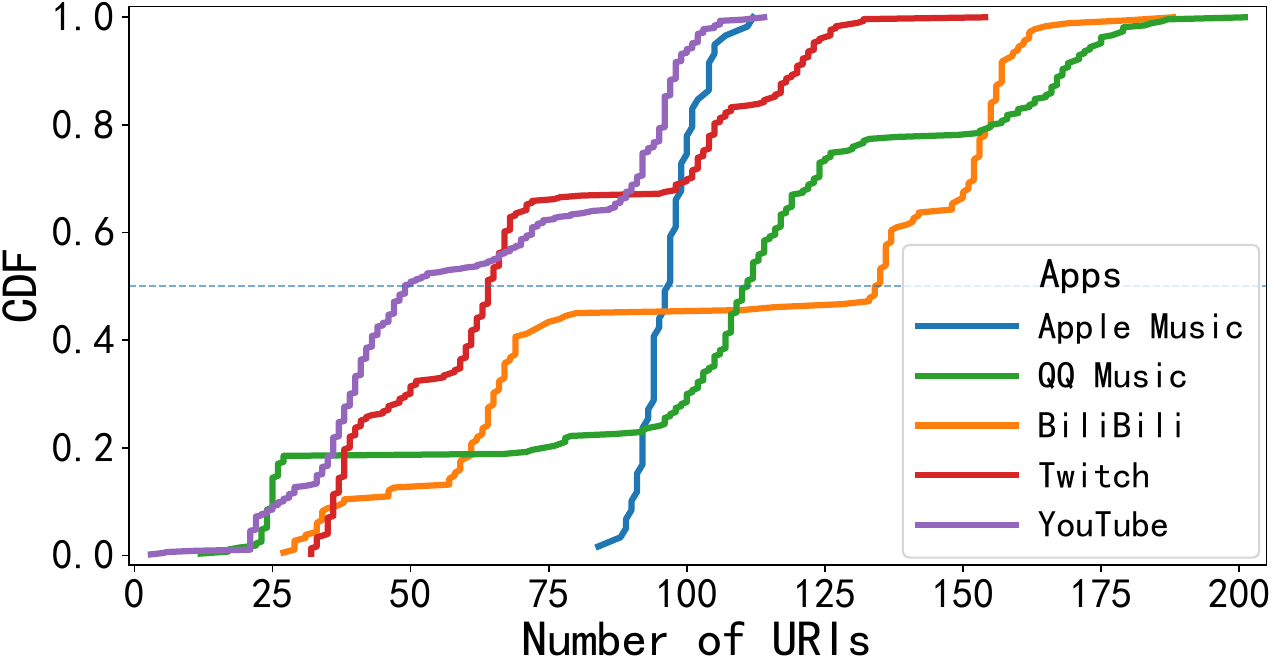}
    \caption{CDF of the number of distinct URIs}
    \label{fig:uri_cdf}
\end{figure}

Moreover, \projname{} does not rely solely on single URI predictions for behavior inference. Instead, it leverages the invocation patterns of multiple URIs to construct a CUM for each behavior during training.
Specifically, during the training phase, the same behavior of applications is repeatedly executed.
By statistically analyzing the sampled data and grouping flows according to their backend services (i.e., domains), we identify the most frequent URI invocation patterns, which are then used to construct CUM, with each branch of CUM corresponding to a specific domain.
As demonstrated in Figure~\ref{fig:uri_cdf}, even a single behavior such as \texttt{Play} may involve dozens to hundreds of distinct URIs, underscoring the structural richness and diversity of backend interactions. By abstracting these recurring patterns into CUMs, \projname{} provides robust and semantically grounded templates that enhance the accuracy and interpretability of fine-grained behavior inference across platforms.

\subsubsection{URI Map-based Fine-Grained Inference}\label{sec:uri_matching}

We then compare the predicted URI map with the pre-built CUMs of candidate behaviors of all potential applications (as multiple applications may pass the coarse-grained test in Section~\ref{sec:coarse}). This comparison integrates both the temporal structure of URI maps and the confidence scores of individual URI predictions to identify the most likely application behavior, thereby determining the originating application.

We adopt a URI map-based matching algorithm (Algorithm~\ref{alg:WeightedLCS}) to estimate the likelihood that a predicted URI map (e.g., $\mathcal{S}$) corresponds to a candidate behavior’s CUM (e.g., $\mathcal{C}$). 
Predicted URIs are first grouped by their destination domains to form domain-specific branches ordered by timestamps. Each branch is then aligned with the corresponding domain subgraph in the CUM using the standard longest common subsequence (LCS) algorithm. 
Let $\mathcal{M}$ denote the set of matched URIs across all branches, and let $\mathcal{P}$ represent the intersection of URIs between the predicted map and the CUM. The similarity score is defined as:
\begin{equation}\label{eq:map_score}
    Score_{map} = \frac{\sum_{u_i \in \mathcal{M}} p_{u_i}} {\sum_{u_j \in \mathcal{P}} p_{u_j} + \lambda(|\mathcal{C}| - |\mathcal{P}|)},  \quad p_{u_i} \geq 0.5
\end{equation}
where $p_{u_i}$ and $p_{u_j}$ denote the confidence scores of URIs in the predicted map. Only matched URIs with confidence scores above 0.5 are considered, since low-confidence predictions may reduce reliability. If multiple instances of the same URI appear in both the predicted map and the CUM, $p_{u_j}$ is set to the maximum confidence value. Here, $|\mathcal{C}|$ and $|\mathcal{P}|$ denote the number of URIs in each set (without duplicates). To penalize incomplete coverage, a term $|\mathcal{C}| - |\mathcal{P}|$ is added to the denominator, controlled by a coverage penalty $\lambda$. Therefore, $Score_{map}$ achieves its maxima (i.e., 1) when all URIs in the CUM are perfectly matched with high confidence.


Next, we finalize fine-grained behaviors and the originating applications using two strategies. (i) \textbf{Segment level.} For each active window of each potential application, we select the application behavior with the maximum score; if no candidate exceeds a preset unseen threshold $\beta$, the segment is temporarily tagged as \textit{unseen}. Note that the choice of the coverage penalty $\lambda$ and the unseen threshold $\beta$ is interdependent, which will be evaluated in Section~\ref{sec:unseen_detection}. (ii) \textbf{Traffic level.} Because coarse-grained filtering (Section~\ref{sec:segments-filter}) may yield overlapping activity windows across applications, we disambiguate by flow attribution. Within each activity window, we first identify the flows that contributed to the URI-map matching process (i.e., those aligned by the LCS algorithm). When a flow is claimed by multiple applications, we compare their $Score_{map}$ values and assign the flow to the application with the highest score, discarding the others. After iterating over all flows, any segments still carrying the \textit{unseen} tag are treated as real unseen cases, then forwarded to the next stage to refine behavior inference. Importantly, this scheme also decouples overlapped bursts generated by simultaneous users across different devices, thereby supporting multi-user scenarios.

\subsubsection{Refinement of Unseen Cases}
\label{sec:unseen-handling}

Three scenarios can lead to unseen cases of interest in \projname{}. (i) Unseen applications, where flows are produced by applications absent from the training set but sharing similar functionalities with at least one trained application. (ii) Unseen platforms, where the same application runs on an unobserved platform such as IoT devices or new operating systems. (iii) Unseen versions, where a major update of the same application introduces noticeable changes in traffic patterns. \projname{} considers only those unseen cases that exhibit similarity to the existing dataset, meaning those that pass both the coarse-grained test in Section~\ref{sec:coarse} and the fine-grained test in Section~\ref{sec:uri_matching}, and ignores traffic from purely unrelated applications, for which meaningful inference is impractical.



We find that the key to refine the inference of open-world unseen cases is to distinguish the shared features (i.e., the shared URIs) and ignore the customized features (i.e., the private URIs). Because private URIs in unseen cases are inherently unknown, their confidence scores bring noise into matching process. We therefore adopt a simple yet effective refinement: discard private URIs in the CUM and compute the matching score in Eq.~\ref{eq:map_score} using only shared URIs. Note that, our objective here is not to estimate the exact proportion of shared URIs, but to suppress noise from private URIs and thus stabilize matching for unseen cases. As a result, we can simply identify the shared URIs from cross-platform applications in the training set.

\comment{
\subsubsection{Application Identification \zyz{?}}
In the initial phase of traffic filtering, it is expected that this module will encounter some inevitable false positives, particularly when dealing with similar applications or other complex scenarios. These false positives arise because different applications may share similar behavior patterns or API request sequences, making it difficult to distinguish them accurately. As a result, a rough classification of applications generating the network traffic is performed. However, this initial classification often needs refinement.

To improve accuracy, we rely on the outcomes of API identification to further refine the application classification. By leveraging the API identification model, we gain insights into the API request patterns over a specific period, including the frequency and sequence of API calls. For network flows that remain uncertain after the initial classification, we integrate the contextual information of these flows, such as their temporal behavior and the relationships between different API requests.

To enhance the precision of application identification, we assess the degree of alignment between the behavior logic graph and the API request list. By comparing the contextual features of the network flow with the structure of the behavior graph, we can more accurately determine which application is responsible for generating the network flow. This method enables us to handle complex cases, such as distinguishing between similar applications, by analyzing the finer details of their behavior patterns.
}

%% file: 6_eval.tex
\section{Evaluation}\label{sec:eval}

We first present an overall evaluation of \projname{} in cross-platform settings in Section~\ref{sec:eval_cross}, including Android, iOS, and PC, before assessing the performance of open-world scalabilities in Section~\ref{sec:eval_scalability}.

\subsection{Dataset Construction}
Since our work is the first to leverage URI-level, platform-agnostic characteristics for cross-platform and open-world traffic fingerprinting, no public dataset is available. We therefore constructed our own dataset. First, we downloaded 500 popular applications (ranked by store listings) from both Google Play and the iOS App Store, respectively, covering 29 categories to ensure reasonable comprehensiveness. Applications requiring credit-card information at registration were excluded to avoid potential legal risks. Second, for desktop, we also collected traffic from the corresponding web or native desktop versions of 500 mobile applications. We used Airtest~\cite{yu2021layout} for cross-platform automated interaction because it relies on computer-vision-based UI manipulation and thus generalizes across platforms. For each application, scripts executed 1,000 random clicks in 50 traffic instances to simulate random user behaviors, where a traffic instance contains 20 clicks. We record UI interactions and manually associate user actions with each decrypted URI according to timestamps, since these traffic instances were collected in a controlled MitM environment. 
In subsequent experiments, we constructed training set based on the collected instances, and trained four models using this dataset: a flow-level similarity model, a logistic flow filtering model, a burst-level URI classification model, and the CUM. For applications available on multiple platforms, each platform-specific version was treated as a distinct application. Experiments were conducted on Android (Google Pixel 7a and Redroid), iOS (iPhone XS Max and iPhone SE), and desktop (virtual machines), with all devices connected to the same network.

\input{Table/eval1.tex}

\subsection{Cross-Platform Performance}\label{sec:eval_cross}

\subsubsection{Performance of Fine-Grained Fingerprinting}
We first evaluate the performance of application and detailed behavior identification under cross-platform conditions.

\noindent\textbf{Experimental Setup.} 
This study uses encrypted traffic collected from Android, iOS, and Windows platforms and compares performance with established benchmarks such as APPScanner~\cite{2016AppScanner} and FOAP~\cite{li2022foap}. For each application, traffic instances are randomly divided into two subsets: 40 for training and 10 for testing. To emulate two simultaneous users on different devices, testing network traces are generated by probabilistically (50\%) merging two random traffic instances with a random time delay of 0–5 seconds~\cite{zungur2020libspector}. Two experiments are conducted: (i) comparing the performance of \projname{} against standard benchmarks, and (ii) extending benchmark methods using our URI-level dataset, which provides finer-grained network characteristics than client-side instrumentation. Note that APPScanner supports only application fingerprinting, and behavior-level fingerprinting results are therefore not reported. Standard evaluation metrics, including precision, recall, and F1-score, are used to assess identification performance.

\noindent\textbf{Results.}
Table~\ref{tab:interface_behavior_comparison} summarizes the experimental results, from which we draw two key observations. (i) \projname{} consistently outperforms the benchmark methods in both application and behavior recognition. In particular, for behavior recognition, \projname{} achieves notable improvements, with average precision increased by around 0.176, recall by 0.534, and F1-score by 0.450. This is not surprising, as \projname{} effectively decouples applications across different platforms and interleaved traffic generated by multiple users and platforms. By contrast, FOAP exhibits substantial performance degradation under these conditions. (ii) Compared with the original APPScanner and FOAP, the extended versions demonstrate significant improvement. This gain arises because using URIs as the unit of analysis largely reduces label noise introduced by client-side features, as URI-level characteristics more directly capture underlying network behaviors.

\subsubsection{Performance of URI Classification}\label{sec:effect_burst}

\input{Table/deltaTime.tex}
We next evaluate the URI classification accuracy.

\noindent \textbf{Experimental Setup.} 
URI classification is considered as an application-specific multi-class classification task, where each flow burst is assigned to a backend URI label. To evaluate the impact of burst segmentation on URI classification performance, we vary the inter-burst gap threshold $\Delta t$ from 50 ms to 5000 ms, which determines how encrypted flows are grouped into coherent bursts. Following the same procedure as before, the traffic instances of each application are randomly split into two subsets: 40 instances for training and 10 instances for testing. It is worth noting that in this experiment we omit the first-stage coarse-grained application filtering and directly apply the corresponding model for URI classification, as predictions produced by an incorrect model would be meaningless.

\noindent \textbf{Results.}
Table~\ref{tab:deltaTime} reports the effect of the inter-burst gap threshold $\Delta t$ on URI classification accuracy. When $\Delta t = 50$ ms, precision remains high (0.868), but the F1-score drops to 0.829 because overly aggressive segmentation often fails to capture a complete URI sequence. Increasing $\Delta t$ to 500 ms improves performance, yielding precision of 0.879, recall of 0.838, and an F1-score of 0.848, as bursts more accurately capture the side-channel features of each URI invocation. Further increasing $\Delta t$ to 2000 ms or 5000 ms leads to declines across all metrics, since larger bursts tend to mix multiple URIs, obscuring boundaries and reducing clarity. These results indicate that $\Delta t = 500$ ms achieves the best trade-off between segmentation granularity and semantic completeness. Importantly, URI-level classification accuracy does not directly bound downstream application- and behavior-recognition performance, as temporal, structure-aware URI maps can further improve accuracy and robustness.


\subsubsection{Impact of URI Maps}
Then we evaluate the impact of URI map-based fine-grained matching algorithm.

\noindent \textbf{Experimental Setup.}  
We constructed experimental scenarios by selecting 10 similar applications from each of three categories for testing: Social Media (Social), Video apps, and Music apps. For the baseline URI-based method, behavior identification is performed based on the proportion of predicted URIs compared with the canonical URI set, without leveraging the confidence scores and logical sequence information employed in the URI map-based approach. For both methods, we measured the false positive rate (FPR) and false negative rate (FNR) in behavior identification across the selected categories.

\input{Table/API_Behavior.tex}

\noindent \textbf{Results.}  
Table~\ref{tab:fnr_fpr} summarizes the results. On average, the URI-based approach produces $\mathrm{FNR}=0.172\pm0.120$ and $\mathrm{FPR}=0.072\pm0.084$. Certain categories, such as \textit{Music}, exhibit higher baseline errors ($\mathrm{FNR}=0.248$, $\mathrm{FPR}=0.122$), highlighting variability across application types. By contrast, aligning predicted URIs with the temporally structured CUM significantly reduces errors to $\mathrm{FNR}=0.037\pm0.046$ and $\mathrm{FPR}=0.012\pm0.015$, corresponding to relative reductions of 78.1\% and 83.1\%, respectively. In addition to lowering error rates, the URI Map also reduces variance across categories, yielding more consistent and robust behavior identification. These findings show that incorporating confidence values and temporal structure in the CUM substantially enhances the accuracy and robustness of fine-grained behavior recognition.

\subsection{Open-World Scalability}\label{sec:eval_scalability}
In real-world deployments, behavior identification systems inevitably encounter unseen usage cases. These can be broadly divided into two categories: (i) entirely irrelevant cases, which can be directly excluded through the first-stage coarse-grained application filtering, without affecting the overall identification performance, and (ii) cases that exhibit partially similar patterns, such as cross-platform migrations, application variants with high functional overlap, or updates to application versions. For the latter category, fine-grained behaviors can still be inferred using the URI Map–based matching algorithm. In this section, we evaluate the scalability of \projname{} under open-world settings.

\subsubsection{Performance of Unseen Case Detection} 
\label{sec:unseen_detection}
We begin by evaluating the ability of \projname{} to accurately identify genuine unseen cases.

\noindent\textbf{Experimental Setup.}
The effectiveness of unseen case detection depends on two key parameters: the penalty coverage $\lambda$ and the unseen threshold $\beta$. To identify parameter values, we perform a grid search to determine the combination that yields the best detection performance. The evaluation is conducted on a dataset of 40 randomly selected applications, with 20 included in training (known) and 20 excluded from training (unseen). As in earlier experiments, we omit the first-stage coarse-grained filtering and directly apply URI classification and URI Map matching to all test applications for unseen case detection. The parameter $\lambda$ is varied over the range [0.2, 2], while $\beta$ is adjusted within the range [0.1, 0.9]. For each setting, F1-scores are computed to quantify detection performance.

\begin{figure}[t]
    \centering
    \includegraphics[width=0.9\linewidth]{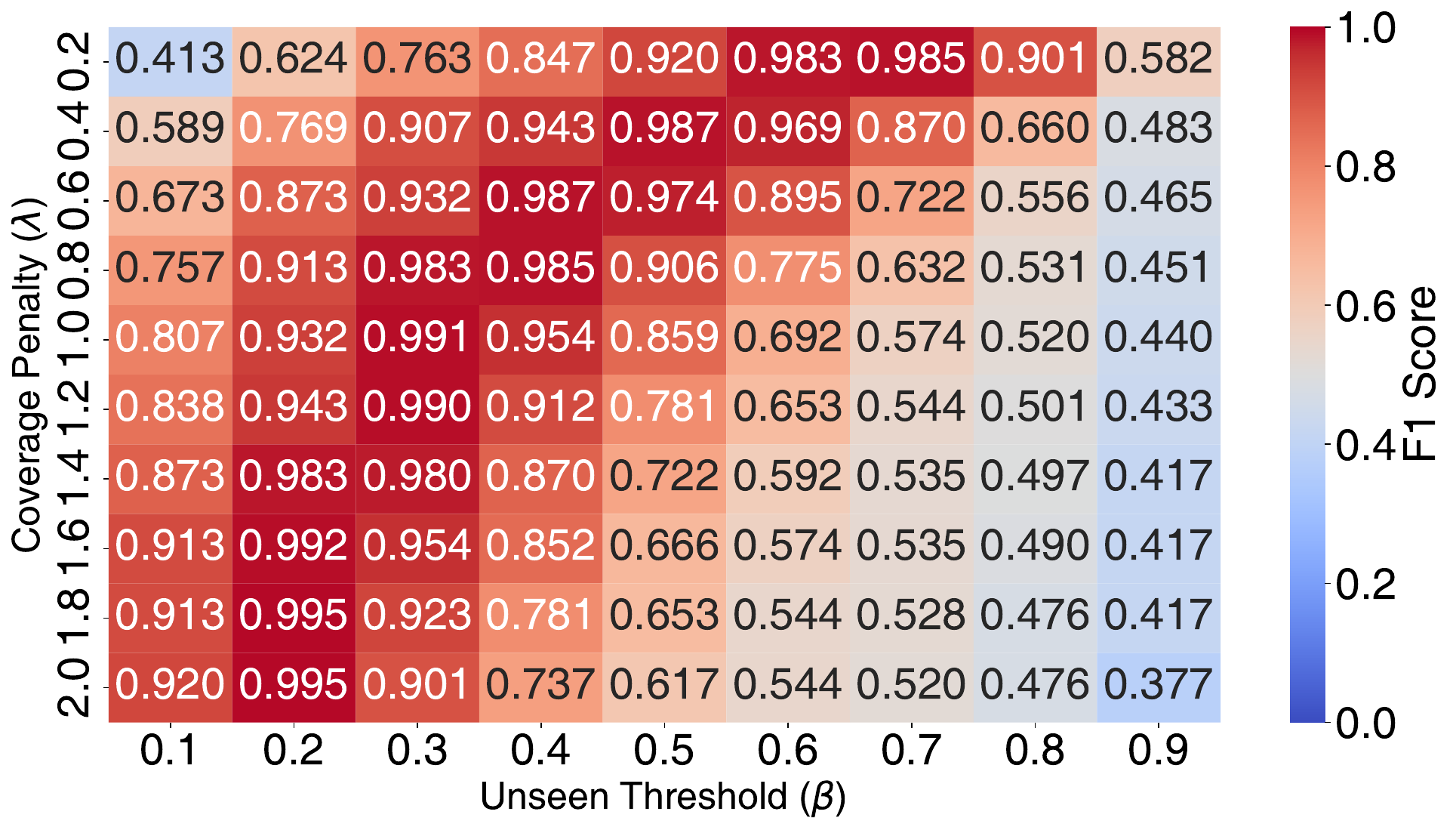}
    \vspace{-10pt}
    \caption{Impact of coverage penalty $\lambda$ and unseen threshold $\beta$ on the F1-score for unseen application detection. Higher values indicate better performance.}
    \label{fig:lambda_heatmap}
\end{figure}

\input{Table/platform.tex}

\input{Table/domain.tex}

\input{Table/eval_temporal.tex}


\noindent\textbf{Results.} Figure~\ref{fig:lambda_heatmap} presents the results. For each value of $\lambda$, as $\beta$ increases, the F1-score initially rises and then declines, reflecting the fact that $\beta$ directly influences the trade-off between false positives and false negatives in unseen case detection. According to Eq.~\ref{eq:map_score}, larger values of $\lambda$ impose heavier penalties on mismatches, thereby reducing URI map matching scores and lowering the effective unseen threshold $\beta$. The choice of $\lambda$ impacts not only unseen case detection but also the accuracy of known behavior matching. To ensure that the average matched similarity $Score_{map}$ for known cases remains above 0.8, thereby improving the robustness of behavior identification, $\lambda$ should be at most 1, as validated on our dataset. Based on these observations, we select $\lambda=1$ and $\beta=0.3$, which jointly maximize the F1-score for unseen case detection. These findings confirm that \projname{} can effectively detect unseen cases by carefully tuning these parameters.

\subsubsection{Unseen-Platform Evaluation} 
We next evaluate performance on traffic from unseen platforms, including previously unobserved systems (e.g., smart TVs) and previously unobserved OS versions (e.g., iOS~17).

\noindent \textbf{Experimental setup.} 
To assess generalization under platform and OS-version shifts, we consider two unseen-target settings: models trained on Android, iOS~15, and Windows are tested on Smart~TV (platform migration) and on iOS~17 (OS-version migration), respectively. We randomly select 20 applications for training, restricting selection to apps available on both Smart~TV and iOS~17 so that source and target sets are well aligned. The pipeline applies coarse application filtering followed by fine-grained URI map matching. We report precision, recall, and $F_{1}$, comparing FOAP, \projname{} without refinement, and the full \projname{} w/ refinement.

\noindent\textbf{Results.} Table~\ref{tab:cross-platform} shows that \projname{} w/ refinement substantially outperforms both FOAP and \projname{} w/o refinement in cross-platform migration and platform version drift scenarios. During testing, all samples passed the initial coarse-grained application filtering stage, which is expected given the deliberately loose filtering criterion. Two key observations emerge from the results. First, even without refinement, \projname{} already surpasses FOAP, as the use of finer-grained URI characteristics effectively eliminates ambiguous endpoints and reduces label noise. Second, the refinement process provides additional improvements by suppressing noise from private URIs and stabilizing matching process in unseen cases.

\subsubsection{Unseen-Application Evaluation} 
Many applications with similar functionalities share third-party SDKs or are maintained by the same service provider, which often results in partial URI overlap. This study evaluates whether \projname{} can reliably distinguish between different applications under such conditions of overlap.

\noindent\textbf{Experimental setup.} We construct the test set by grouping applications into three functional categories: Social Media (Social), Video, and Music. For each category, 20 applications are randomly selected, comprising 10 known and 10 unseen apps. We evaluate three methods—FOAP, \projname{} w/o refinement, and the full \projname{}—and report precision, recall, and $F_{1}$ for behavior identification in each category.

\noindent\textbf{Results.} Table~\ref{tab:similar-unseen} shows that \projname{} consistently outperforms both FOAP and \projname{} w/o refinement across all categories, with average precision improvements of approximately 0.34 and 0.294, recall gains of 0.268 and 0.201, and F1-score increases of 0.319 and 0.307, respectively. The results indicate that \projname{} w/o refinement is susceptible to noise from private URIs, which cause mismatches and degrade performance. By contrast, the refined \projname{} effectively suppresses such noise during matching, thereby enhancing accuracy and robustness. These findings highlight the importance of distinguishing shared features from private ones in unseen applications, as this separation is crucial for reliable behavior identification in open-world settings.



\subsubsection{Unseen-Version Evaluation} 
Prior work has often struggled with version shifts. Next,  we evaluate \projname{} over nearly one year of application releases to assess its ability to handle major version updates.

\noindent\textbf{Experimental setup.} 
We select ten applications from the Video, Social Media, and Music categories, as listed in Appendix~\ref{sec:appendix_version}, with three representative behaviors per application. For each application, encrypted traffic is collected at two release points—October 2024 and August 2025 (an 11-month interval)—using identical data collection scripts. Models are trained on the October 2024 data and evaluated on the August 2025 data without additional tuning. We report precision, recall, and F1-score of behavior identification.

\noindent\textbf{Results.} 
Table~\ref{tab:api-change-comparison} shows that \projname{} consistently outperforms both FOAP and \projname{} w/o refinement across all categories, with average precision gains of approximately 0.405 and 0.327, recall gains of 0.355 and 0.240, and F1-score improvements of 0.417 and 0.265, respectively. These results demonstrate the effectiveness of \projname{} in handling major version updates without additional tuning. Although application updates often modify network behaviors, the underlying backend services associated with core functionalities—and their corresponding URIs—typically remain stable for backward compatibility. This stability enables \projname{} to leverage shared URIs as reliable anchors for behavior identification, thereby maintaining robustness as applications evolve.

\comment{
\subsection{Cross-Dataset Evaluation}
\zyk{In this experiment, we evaluate the impact of the sample data automatically collected by \projname{} from scripts on the identification of real human operations. }

\input{Table/RHBehavior.tex}

\noindent\textbf{Experimental setup.} \zyk{To assess generalization from the randomized traces, we additionally built a human-operated dataset: five volunteers manually interacted with 100 applications, producing 500 traffic instances with realistic usage patterns on the same three platforms. We construct a human generated dataset by randomly sampling 100 applications from a pool of 1,000 and recruiting five volunteers. Each volunteer operates all 100 apps for about five minutes per session, producing 5$\times$100 traffic instances. We evaluate three transfer settings: A$\rightarrow$A trains and tests \projname{} on Airtest generated traffic from the same domain; A$\rightarrow$H trains on Airtest traffic and tests on human traffic; AH$\rightarrow$H augments the Airtest training set with human traffic from four volunteers and tests on the remaining volunteer. We rotate the held out volunteer and report the average for statistical reliability. Models are trained only on the designated source data and evaluated on the target without additional tuning. We report behavior level precision, recall, and F1.}

\noindent \textbf{Results.} \zyk{The results in Table~\ref{tab:RHbehavior} show high identification accuracy in \textit{A$\rightarrow$A} and \textit{A$\rightarrow$H}, indicating that our Airtest based collection effectively reproduces human behavior. In \textit{A+H$\rightarrow$H}, \projname{} attains perfect precision, recall, and F1. Compared with FOAP, \projname{} maintains stable performance, whereas FOAP’s F1 drops in \textit{A$\rightarrow$H} due to context dependent inference. This experiment highlights an efficient training data strategy that combines automated scripts for large scale collection with human sessions to capture real world operational logic, thereby reducing false negatives and improving recall for \projname{}.}
}

%% file: Table/eval1.tex
\begin{table*}[t]
\centering
\small
\caption{Performance comparison of app and behavior identification under cross-platform settings (mean~$\pm$~standard deviation).}
\vspace{0.5em} 
\label{tab:interface_behavior_comparison}
\renewcommand{\arraystretch}{0.75}
\resizebox{\textwidth}{!}{
\begin{tabular}{lcccccc}
\toprule
\multirow{2}{*}{\textbf{Method}} 
& \multicolumn{3}{c}{\textbf{App Recognition}} 
& \multicolumn{3}{c}{\textbf{Behavior Recognition}} \\
\cmidrule(lr){2-4} \cmidrule(lr){5-7}
& \textbf{Precision} & \textbf{Recall} & \textbf{F1-Score} 
& \textbf{Precision} & \textbf{Recall} & \textbf{F1-Score} \\
\midrule
APPScanner 
& 0.806 $\pm$ 0.029 & 0.717 $\pm$ 0.056 & 0.691 $\pm$ 0.070 
& ----    & ----   & ----    \\
APPScanner (extended)
& 0.927 $\pm$ 0.019 & 0.920 $\pm$ 0.025 & 0.920 $\pm$ 0.025 
& ----   & ----    & ----    \\
FOAP 
& 0.837 $\pm$ 0.099 & 0.896 $\pm$ 0.069 & 0.859 $\pm$ 0.054
& 0.794 $\pm$ 0.072 & 0.436 $\pm$ 0.239 & 0.515 $\pm$ 0.210 \\
FOAP (extended)
& 0.978 $\pm$ 0.029 & 0.939 $\pm$ 0.046 & 0.957 $\pm$ 0.028 
& 0.879 $\pm$ 0.117 & 0.881 $\pm$ 0.113 & 0.877 $\pm$ 0.117 \\

\projname{}
& \textbf{0.996} $\pm$ \textbf{0.005} & \textbf{0.940} $\pm$ \textbf{0.142} & \textbf{0.977} $\pm$ \textbf{0.107} 
& \textbf{0.970} $\pm$ \textbf{0.035} & \textbf{0.970} $\pm$ \textbf{0.035} & \textbf{0.965} $\pm$ \textbf{0.043} \\
\bottomrule
\end{tabular}
}
\end{table*}

%% file: Table/deltaTime.tex


\begin{table}[t]
\centering
\caption{Impact of inter-burst  $\Delta t$ on URI identification.}
\vspace{0.5em} 
\label{tab:deltaTime}
\small
\resizebox{\linewidth}{!}{
\begin{tabular}{cccc}
\toprule
\textbf{$\Delta t$ (ms) }& \textbf{Precision} & \textbf{Recall} & \textbf{F1-Score} \\
\midrule
50    & 0.868$\pm$0.036 & 0.812$\pm$0.036 & 0.829$\pm$0.036 \\
500   & \textbf{0.879$\pm$0.091} & \textbf{0.838$\pm$0.093} & \textbf{0.848$\pm$0.092} \\
2000  & 0.829$\pm$0.096 & 0.814$\pm$0.097 & 0.814$\pm$0.094 \\
5000  & 0.811$\pm$0.123 & 0.798$\pm$0.125 & 0.796$\pm$0.120 \\
\bottomrule
\end{tabular}
}
\end{table}

%% file: Table/API_Behavior.tex
\begin{table}[t]
\centering
\small
\setlength{\tabcolsep}{8pt}
\caption{FNR/FPR for behavior identification.}
\renewcommand{\arraystretch}{0.75}
\vspace{0.5em}
\label{tab:fnr_fpr}\resizebox{\linewidth}{!}{
\begin{tabular}{ccccc}
\toprule
\multirow{2}{*}{\textbf{Category}} & \multicolumn{2}{c}{\textbf{URI-based}} & \multicolumn{2}{c}{\textbf{URI Map-based}} \\
\cmidrule(lr){2-3}\cmidrule(lr){4-5}
& \textbf{FNR} & \textbf{FPR} & \textbf{FNR} & \textbf{FPR} \\
\midrule
Social         & 0.134±0.103 & 0.046±0.025 & 0.039±0.056 & 0.010±0.014 \\
Video          & 0.133±0.114 & 0.049±0.049 & 0.028±0.033 & 0.012±0.014 \\
Music          & 0.248±0.105 & 0.122±0.123 & 0.045±0.045 & 0.015±0.015 \\
\midrule
\textbf{Average} & \textbf{0.172±0.120} & \textbf{0.072±0.084} & \textbf{0.037±0.046} & \textbf{0.012±0.015} \\
\bottomrule
\end{tabular}
}
\end{table}

%% file: Table/platform.tex

\begin{table*}[t]
\centering
\small
\caption{Fine-grained behavior identification performance for unseen platforms (mean $\pm$ standard deviation).}
\vspace{0.5em}
\label{tab:cross-platform}
\small
\resizebox{\linewidth}{!}{
\begin{tabular}{lccccccccc}
\toprule
\multirow{2}{*}{\textbf{Platforms Shift}} & \multicolumn{3}{c}{\textbf{FOAP}} & \multicolumn{3}{c}{\textbf{\projname{} (w/o Refinement)}} & \multicolumn{3}{c}{\textbf{\projname{}}} \\ \cline{2-10} 

                    & \textbf{Precision}    & \textbf{Recall}    & \textbf{F1-Score}     & \textbf{Precision}          & \textbf{Recall}         & \textbf{F1-Score}          & \textbf{Precision}    & \textbf{Recall}    & \textbf{F1-Score} \\ \midrule
Android + iOS15 + Windows $\rightarrow$ SmartTV & 0.512$\pm$0.119    & 0.490$\pm$0.122 & 0.501$\pm$0.120   & 0.614$\pm$0.117          & 0.589$\pm$0.122       & 0.601$\pm$0.120         & 0.802$\pm$0.116    & 0.724$\pm$0.123 & 0.747$\pm$0.118   \\
Android + iOS15 + Windows $\rightarrow$ iOS17 & 0.524$\pm$0.117    & 0.511$\pm$0.119 & 0.517$\pm$0.118   & 0.621$\pm$0.121          & 0.609$\pm$0.118       & 0.615$\pm$0.119         & 0.833$\pm$0.115    & 0.730$\pm$0.121 & 0.744$\pm$0.118   \\ 
\bottomrule
    \end{tabular}}
\end{table*}

%% file: Table/domain.tex
\begin{table*}[t]
\small
\centering
\caption{Fine-grained behavior identification performance for unseen applications (mean $\pm$ standard deviation).}
\vspace{0.5em} 
\label{tab:similar-unseen}

\renewcommand{\arraystretch}{1} 
\resizebox{\textwidth}{!}{
\begin{tabular}{cccccccccc} 
\toprule
\multirow{2}{*}{\textbf{Category}} & \multicolumn{3}{c}{\textbf{FOAP}} & \multicolumn{3}{c}{\textbf{\projname{} (w/o Refinement)}} & \multicolumn{3}{c}{\textbf{\projname{}}} \\ \cline{2-10} 
                     & \textbf{Precision}    & \textbf{Recall}    & \textbf{F1-Score}     & \textbf{Precision}          & \textbf{Recall}         & \textbf{F1-Score}          & \textbf{Precision}    & \textbf{Recall}    & \textbf{F1-Score} \\ \midrule

\textbf{Video}                & 0.528±0.011                            & 0.524±0.008                         & 0.507±0.005                           & 0.421±0.171                            & 0.536±0.036                         & 0.440±0.106                           & 0.834±0.118                            & 0.742±0.188                         & 0.765±0.170                           \\
\textbf{Social}               & 0.356±0.002                            & 0.355±0.002                         & 0.353±0.004                           & 0.477±0.156                            & 0.468±0.008                         & 0.406±0.029                           & 0.834±0.118                            & 0.755±0.186                         & 0.772±0.172                           \\
\textbf{Music}                & 0.473±0.019                            & 0.472±0.020                         & 0.471±0.020                           & 0.595±0.119                            & 0.548±0.071                         & 0.523±0.048                           & 0.708±0.093                            & 0.657±0.094                         & 0.672±0.092                           \\ \hline
\textbf{Average}              & \multicolumn{1}{c}{0.452±0.011}        & \multicolumn{1}{c}{0.450±0.010}     & \multicolumn{1}{c}{0.444±0.010}       & \multicolumn{1}{c}{0.498±0.149}        & \multicolumn{1}{c}{0.517±0.038}     & \multicolumn{1}{c}{0.456±0.061}       & \multicolumn{1}{c}{0.792±0.110}        & \multicolumn{1}{c}{0.718±0.156}     & \multicolumn{1}{c}{0.736±0.145}       \\ \bottomrule
\end{tabular}
}
\end{table*}

%% file: Table/eval_temporal.tex
\begin{table*}[t]
\small
\centering
\caption{Fine-grained behavior identification performance for unseen application versions (mean $\pm$ standard deviation).}
\vspace{0.5em}
\label{tab:api-change-comparison}

\renewcommand{\arraystretch}{1}
\resizebox{\textwidth}{!}{
\begin{tabular}{cccccccccc} 
\toprule
\multirow{2}{*}{\textbf{Category}} & \multicolumn{3}{c}{\textbf{FOAP}} & \multicolumn{3}{c}{\textbf{\projname{} (w/o Refinement)}} & \multicolumn{3}{c}{\textbf{\projname{} }} \\ \cline{2-10} 

                    & \textbf{Precision}    & \textbf{Recall}    & \textbf{F1-Score}     & \textbf{Precision}          & \textbf{Recall}         & \textbf{F1-Score}          & \textbf{Precision}    & \textbf{Recall}    & \textbf{F1-Score} \\ \midrule
\textbf{Video}       & 0.448 $\pm$ 0.131 & 0.216 $\pm$ 0.135 & 0.279 $\pm$ 0.156  & 0.659 $\pm$ 0.036    & 0.563 $\pm$ 0.024    & 0.601 $\pm$ 0.013   & 0.833 $\pm$ 0.036   & 0.710 $\pm$ 0.010   & 0.736 $\pm$ 0.074   \\
\textbf{Social}      & 0.397 $\pm$ 0.119 & 0.322 $\pm$ 0.157 & 0.342 $\pm$ 0.151 & 0.334 $\pm$ 0.134    & 0.321 $\pm$ 0.012    & 0.290 $\pm$ 0.040   & 0.810 $\pm$ 0.050   & 0.655 $\pm$ 0.045   & 0.691 $\pm$ 0.109   \\
\textbf{Music}       & 0.415 $\pm$ 0.020 & 0.434 $\pm$ 0.058 & 0.278 $\pm$ 0.050 & 0.500 $\pm$ 0.110    & 0.434 $\pm$ 0.019    & 0.464 $\pm$ 0.011   & 0.833 $\pm$ 0.067   & 0.672 $\pm$ 0.028   & 0.723 $\pm$ 0.077   \\ \hline
\textbf{Average}     & 0.420 $\pm$ 0.090 & 0.324 $\pm$ 0.117 & 0.300 $\pm$ 0.119 & 0.498 $\pm$ 0.093    & 0.439 $\pm$ 0.018    & 0.452 $\pm$ 0.021   & 0.825 $\pm$ 0.051   & 0.679 $\pm$ 0.028   & 0.717 $\pm$ 0.087   \\
\bottomrule
\end{tabular}
}
\end{table*}

%% file: Table/RHBehavior.tex
\begin{table}[ht]
    \small
\caption{Behavior identification on the cross-dataset experimental settings (mean$\pm$standard deviation).}
\label{tab:RHbehavior}

\vspace{0.5em}
\centering
\resizebox{\linewidth}{!}{
\begin{tabular}{cccccc}
\hline
\textbf{Transfer Setting} & \textbf{Precision} & \textbf{Recall} & \textbf{F1-Score} \\ 
\hline
A $\rightarrow$ A & 0.933$\pm$0.088 & 0.956$\pm$0.055 & 0.927$\pm$0.099 \\
A $\rightarrow$ H      & 0.932$\pm$0.086 & 0.956$\pm$0.054 & 0.927$\pm$0.094 \\
AH $\rightarrow$ H    & 0.973$\pm$0.053 & 0.955$\pm$0.088 & 0.950$\pm$0.100 \\
\hline
\end{tabular}
}
\end{table}


%% file: 3_related.tex
\section{Discussion}\label{sec:discussion}

\textbf{Countermeasures.}  
The fundamental strategy to counter traffic analysis by \projname{} is to disrupt the side-channel signals—such as packet size, direction, and timing—that enable behavior inference. By obfuscating these characteristics, \projname{} can no longer reliably map encrypted flows to specific URI invocations. Several approaches may be employed, including packet padding~\cite{yu2012predicted}, randomized delays~\cite{hogan2022shortor}, and traffic transformation techniques~\cite{bocovich2024snowflake}. However, these countermeasures typically introduce additional overhead and may degrade system efficiency.

\noindent \textbf{Limitations.}  
Despite its effectiveness, \projname{} has certain limitations. (i) It is not applicable in Tor-like environments~\cite{karunanayake2021anonymisation}, where fixed-size cell segmentation and multiplexed scheduling obscure flow semantics, preventing application differentiation. Addressing this challenge may require leveraging Tor-specific features or developing advanced models, which we leave for future work. (ii) Some applications adopt strong anti-MitM defenses, such as certificate pinning or advanced verification mechanisms, which complicate the automated deployment of certificate bypass techniques. In these cases, manual intervention remains necessary~\cite{pourali2024racing}.


\section{Related Work}\label{sec:relatedwork}

Traffic encryption has become ubiquitous, yet recent work shows that network traffic still reveals fine-grained user behaviors beyond coarse app identification. Fine-grained traffic analysis aims to extract such detailed insights by inferring application-layer activities~\cite{feng2025unmasking,shen2020fine} (e.g., distinguishing a Twitter post from a Twitter read) rather than just labeling the application~\cite{saltaformaggio2016eavesdropping,Wei2012ProfileDroid}. Researchers are leveraging semantic flow modeling~\cite{luo2024analyzing} and deep learning~\cite{montieri2021packet} to pinpoint subtle in-app actions or states from encrypted traces. Other techniques combine network patterns with application semantics to improve granularity. FOAP~\cite{li2022foap} infers method-level user actions (tied to UI components) by correlating encrypted flow segments with platform-specific code-level entry points,  thereby capturing sensitive operations that generic app classifiers would otherwise miss.
Another frontier in traffic analysis is achieving robustness across platforms and application changes~\cite{taylor2017robust,mariotte2024channel}, as models trained under closed-world assumptions often falter when confronted with new or evolving apps, different device types, or cross-domain deployments. To address this, researchers have explored transfer learning~\cite{andreoletti2019network,gioacchini2023cross} and domain adaptation techniques~\cite{tobiyama2020large,cui2025trafficllm} to handle unseen traffic patterns. Such solutions still suffer from substantial network noise and platform-specific biases. 

Consequently, the challenge of extracting platform-agnostic characteristics that enable fine-grained behavior recognition in cross-platform and open-world environments still remains.

%% file: 8_conclusion.tex
\section{Conclusion}\label{sec:conclusion}
This paper presents \projname{}, novel server-side, URI-based framework for fine-grained and scalable encrypted traffic fingerprinting across heterogeneous platforms. Comprehensive experiments demonstrate that \projname{} improves the F1-score of fine-grained behavior recognition by 45\% over state-of-the-art baselines in cross-platform settings with interleaved traffic. Moreover, by applying a refinement strategy that excludes application- or platform-specific private features from unseen cases, \projname{} achieves accurate and robust fine-grained inference for unseen applications, platforms, and versions. This capability is largely absent from prior work, demonstrating strong scalability of \projname{}.

%% file: Table/BehaviorMap.tex
\begin{algorithm}[t]
\caption{URI Map-based Matching Algorithm}
\label{alg:WeightedLCS}
\begin{algorithmic}[1]
\Require Predicted sequence $\mathcal{S}=\{(s_i,p_i)\}_{i=1}^m$; candidate set $\mathcal{B}$; domain-partitioned CUM  $\{T_b^{(d)}\}$ for each domain $d$; penalty $\lambda$
\Ensure $\hat{b}$: predicted behavior label
\State $\tau \gets 0.5$
\For{each $b \in \mathcal{B}$}
  \State $\mathcal{C} \gets \bigcup_{d}\mathrm{set}\!\big(T_b^{(d)}\big)$
  \State $p_u \gets \max\{\,p_i \mid s_i=u,\ (s_i,p_i)\in\mathcal{S}\,\}$ for all $u$ in $\mathcal{S}$
  \State $\mathcal{P} \gets \mathcal{C} \cap \{\,u : p_u\ \text{is defined}\,\}$
  \State $\mathcal{M} \gets \emptyset$
  \For{each domain $d$}
    \State $\mathcal{M}^{(d)} \gets \textsc{LCSMatch}\big(\mathcal{S}^{(d)},\,T_b^{(d)};\,\tau\big)$
    \State $\mathcal{M} \gets \mathcal{M} \cup \mathcal{M}^{(d)}$ 
  \EndFor
     \State $w^{(d)}=\sum_{u_i\in \mathcal{M}^{(d)}} p_{u_i}$ with $p_{u_i}\ge 0.5$
  \State $den \gets \sum_{u_j\in \mathcal{P}} p_{u_j}\;+\;\lambda\cdot\big(|\mathcal{C}|-|\mathcal{P}|\big)$
  \State $Score_{\text{map}}(b) \gets  w^{(d)}/den $
\EndFor
\State \Return $\hat{b} \gets \arg\max_{b\in\mathcal{B}} Score_{\text{map}}(b)$
\end{algorithmic}
\end{algorithm}